\documentclass[preprint2]{aastex}

\shorttitle{LBQS and FIRST}
\shortauthors{Hewett, Foltz and Chaffee}

\begin{document}

\title{THE LARGE BRIGHT QUASAR SURVEY. VII. THE LBQS AND FIRST 
SURVEYS\altaffilmark{1,2}
\altaffiltext{1}{Observations reported here were obtained in part at
the MMT Observatory, a joint facility of the Smithsonian Institution
and the University of Arizona} \altaffiltext{2}{Some of the data presented
herein were obtained at the W.M. Keck Observatory, which is operated as
a scientific partnership among the California Institute of Technology,
the University of California and the National Aeronautics and Space
Administration.  The Observatory was made possible by the generous
financial support of the W.M. Keck Foundation}}

\author{Paul C. Hewett}
\affil{Institute of Astronomy, Madingley Rd, Cambridge, CB3 0HA, United Kingdom}
\email{phewett@ast.cam.ac.uk}

\author{Craig B. Foltz}
\affil{MMT Observatory, University of Arizona, Tucson, AZ 85721}
\email{cfoltz@as.arizona.edu}

\and

\author{Frederic H. Chaffee}
\affil{W. M. Keck Observatory, 65--1120 Mamalahoa Highway, Kamuela, HI 96743}
\email{fchaffee@keck.hawaii.edu}

\begin{abstract}

The source catalogue for the Large Bright Quasar Survey (LBQS) and the
FIRST Survey are compared in their regions of overlap to assess the
efficiency of the LBQS selection algorithms. In the 270 deg$^2$ common
to both surveys the LBQS contains $\simeq 100\,000$ stellar and $\simeq
40\,000$ non--stellar objects, while the FIRST catalogue contains
$\simeq 25\,000$ sources. Cross-correlation of these lists yields 67
positional coincidences between known LBQS quasars and FIRST sources
and an additional 19 stellar and 149 non-stellar positional
coincidences with the radio catalogue.  Spectroscopy of all the stellar
matches and two--thirds of the non--stellar matches using the Keck I
telescope and the 6.5--m MMT produces eight new quasars. One BL Lac
object, previously misclassified during the LBQS survey is also
identified.  The straightforward fractional incompleteness of the LBQS
determined from this sample is $13\pm4\%$, in good agreement with the
published estimate of $10\%$.  Furthermore, four of the nine new
objects have redshifts in the range $z=0.7-1.0$, a redshift region
where the LBQS selection is known to have decreased efficiency.  The
distributions of the ratio of radio--to--optical power, apparent
magnitude and spectroscopic properties for the new objects are
consistent with those of the 67 LBQS--FIRST objects previously known.
The consistency of the optical and radio properties of the new objects
with those of the known quasars thus supports the conclusion that no
new population of objects, constituting more than $\simeq 7\%$ of
quasars detected by FIRST, has eluded the LBQS optical selection
techniques.  The percentage of radio--detected quasars in the LBQS
catalogue is found to be $12\pm2\%$, considerably smaller than the
value of $25\%$ advocated by White et al. (2000) based on the First
Bright Quasar Survey (FBQS). Apparent differences in the form of the
number--redshift relation for the LBQS and FBQS samples are shown to
arise in large part from the very different optical passbands used in
the compilation of the surveys.

\end{abstract}


\section{INTRODUCTION}

The last decade has seen dramatic advances in our ability to compile
large samples of quasars according to well--defined quantitative
criteria (e.g. Stocke et al. 1991, Schneider, Schmidt and Gunn 1994,
Hewett, Foltz and Chaffee 1995). However, progress in addressing a
number of longstanding questions concerning the statistical properties
of the quasar population continues to be limited by the possibility
that unquantified selection effects bias the intrinsic distribution of
properties inferred from the observations.

A particularly important recent advance for studies of the quasar
population has come about through the  availability of the FIRST radio
survey (Becker, White \& Helfand 1995). The combination of the faint
limiting flux density, $1\,$mJy, at $1400\,$MHz, high--precision
astrometry and extensive sky coverage has enabled a well--defined
sample of $\sim 1000$ optically bright quasars detected at radio
wavelengths to be compiled for the first time (White et al.  2000).
Initial investigations have revealed the presence of some Broad
Absorption Line (BAL) quasars in the FIRST sample (Becker et al.
2000), apparently contradicting one of the few established differences
between the properties of quasars detected at radio wavelengths and
those of the quasar population as a whole (Stocke et al. 1992).
Quantitative comparison of the spectral energy distributions and the
evolutionary behaviour of the FIRST--quasars to results derived from
optically--selected samples of bright quasars, such as the Large Bright
Quasar Survey (LBQS), offer the prospect of better understanding issues such
as the nature of BALs and the physical origin of the very broad spread
of radio to optical luminosity exhibited within the quasar population.
However, concern still exists that the presence of unrecognised
selection effects in optical samples (e.g.  White et al.  2000; \S 5.3)
will preclude a rigorous intercomparison.

The recent extension of the FIRST radio survey to areas of sky on the
celestial equator provides an overlap with approximately half the area
covered by the LBQS, the largest published
survey of apparently bright optically--selected quasars. It is possible
therefore to perform a direct empirical comparison of the two.
In this paper we present a spectroscopic investigation of the optical
sources that both fall within the LBQS flux--limits and coincide with
radio sources in the FIRST catalogue over an area of $270\,$deg$^2$.

Section 2 contains a brief overview of the main properties of the LBQS
and the FIRST surveys together with the results of the
cross--correlation of the FIRST source catalogue with the flux--limited
optical catalogue that forms the basis for the LBQS quasar search.
Section 3 presents the results of spectroscopic follow--up of those
FIRST sources that pair to optical sources and which had not previously
been investigated as part of the LBQS. The census of new objects
identified during the spectroscopic program that satisfy the LBQS
redshift and flux limits is presented in \S 4. The properties of the
new quasars and an assessment of the effectiveness of the LBQS are
discussed in \S 5. A discussion of the implications of the results of
the LBQS--FIRST comparison for understanding the apparent differences
in the statistical properties of the LBQS and the First Bright Quasar
Survey (FBQS) (White et al. 2000) is presented in \S 6, prior to
a summary of the main conclusions of the investigation given in \S 7.

\section{THE CATALOGUE CROSS--CORRELATION}

\subsection{The Large Bright Quasar Survey}

The LBQS is the largest published survey of optically selected
quasars at bright, $m_B < 19$, apparent magnitudes. A full description
of the survey is given in Hewett, Foltz and Chaffee (1995; Paper VI)
along with positions, magnitudes and redshifts for 1055 quasars in an
effective area of $453.8\,$deg$^2$ over a total area of $537\,$deg$^2$
of sky. The source material for the survey consisted of Automated Plate
Measuring (APM) machine scans of United Kingdom Schmidt Telescope (UKST)
direct and objective--prism plates.  The application of a number of
search algorithms to the low--resolution, objective--prism spectra of
objects in the flux--limited catalogue resulted in the detection of
quasars over an extended redshift range, $0.2 \le z \le 3.4$, that
possess a broad range of spectral energy distributions.  Aspects of the
effectiveness of the LBQS are discussed in \S 5.4 of Paper VI, where it
is concluded that, for objects satisfying the broad--band flux limits
for inclusion in the survey, no known class of quasar spectral energy
distribution would escape detection. However, a significant number of
candidate objects remain unobserved, and it is estimated that the survey
contains $\simeq 90\%$ of the total number of quasars satisfying the
broad--band flux limits.  Attention is also drawn to the likely
systematic incompleteness in the survey for quasars with redshifts $z
\sim 0.8$, where the form of the typical quasar spectral energy
distribution results in a relatively red objective--prism spectrum
lacking particularly strong emission lines or continuum breaks.

Since the publication of Paper VI, observations of further candidate
objects have resulted in the identification of a small number of
additional quasars.  Similarly, observations by other workers of a
small number of objects among the unobserved LBQS candidates have
produced additional quasars. The latter group include several objects
discussed in Paper VI that were not included in the LBQS catalogue due
to book--keeping errors and one erroneous classification resulting from
a ``private communication''. Table 1 provides coordinates,
apparent magnitudes and redshifts of 12 objects that should thus be
included with the objects from Table 4 of Paper VI in any analysis of
the LBQS quasar catalogue. Optical spectra of 9 of the objects in Table
1 are shown in Figure 1.  The extended sample of 1067 quasars, together
with the 8 ``unknown'' objects of Table 6 of Paper VI, which are
potential BL Lacs, form the basis of the comparison with the FIRST
survey described in this paper, although the results and conclusions
are essentially identical if the original catalogue of 1055 quasars
from Paper VI is employed.

The published celestial coordinates of LBQS quasars are accurate to
$\simeq 1\,$arcsec in each ordinate. A significant contribution to the
uncertainty arises from the difficulty in defining the celestial
reference frame to which the APM--measured X--Y centroids of objects on
the photographic plates may be tied. The coordinates published in Paper
VI are based on a reference frame defined by the grid of several
hundred PPM (Roser and Bastian 1988) stars within each UKST field. A
significant improvement over the frame defined using the PPM stars is
now possible through the use of the larger number of stars to fainter
magnitudes that are included in the TYCHO--2 catalogue (H{\o}g et al.
2000).  Employing the celestial coordinates and proper motions of the
$\sim 700$ TYCHO--2 stars within each LBQS UKST field, J2000.0
positions have been rederived for all sources in the APM flux--limited
catalogues from which the LBQS quasars were identified. The resulting
positions show greatly reduced systematic errors as a function of
position on the plates. The root--mean--square positional differences
between the catalogued TYCHO--2 star postions and the APM--positions is
typically $\simeq 0.25\,$arcsec. In practice, the presence of factors
such as nearby fainter images and a variety of emulsion/plate flaws
mean that a significant number of ``outliers'' exist when pairing a
large catalogue of APM sources to a reference catalogue, such as FIRST,
so a matching radius of $\sim 2\,$arcsec is appropriate (see \S 2.3).

\subsection{Faint Images of the Radio Sky at Twenty--centimetres}

The instigation of the FIRST radio survey (Becker et al. 1995),
reaching a faint limiting flux density, $1\,$mJy, at $1400\,$MHz,
coupled with the sky coverage of thousands of square degrees, has
enabled a number of observational investigations not hitherto possible
to be undertaken.  In particular, the survey sources include many
quasars, not just the small fraction of the population detectable in
radio surveys with much higher limiting flux densities. Given the very
accurate celestial positions ($< 0.2\,$arcsec) of the FIRST sources
(White et al.  1997), identifying a large well--defined sample of
optically bright quasars detected at radio wavelengths comparable in
size to the largest existing samples of bright, optically--selected
quasars has at last proved viable (White et al.  2000).

The availability, in mid--1999,  of the FIRST survey catalogue over a
substantial area of sky in a strip at $0^{\circ}$ declination provides
a significant overlap in sky coverage with the LBQS. The FIRST
catalogue used in this paper is that available at 1999 July 21
containing $549\,707$ sources from FIRST observations made over the
period 1993--1998. The sky coverage includes an area of $270\,$deg$^2$
coincident with the survey area of the LBQS, and includes 566 LBQS
quasars and 4 LBQS ``unknowns''.

\subsection{The LBQS quasars and the FIRST Catalogue}

An empirical verification of the appropriate radius to employ for
matching the LBQS and FIRST catalogues was made using the existing LBQS
quasars and unknown objects. Of the 570 LBQS quasars and unknowns
within the area of the FIRST catalogue, 72 LBQS sources possess one or
more FIRST sources within a $10\,$arcsec radius. Sixty--seven of these
have a smallest optical to radio source separations of $\le
2.0\,$arcsec. The large number of optical--radio coincidences within
$\le 2.0\,$arcsec accords with expectations based on the astrometric
accuracy of the catalogues.  The predicted number of such chance
coincidences is $<<1$ and all 67 associations almost certainly reflect
physical association.  In addition, 2 LBQS quasars (B1012+0213,
B1313-0142)\footnote{To avoid potential confusion we employ the
original LBQS object designations derived from the B1950.0 coordinates} lie
along lines joining two catalogued FIRST sources but the closest
optical to radio source separation is $> 2.0\,$arcsec. One LBQS source
(B1319+0039) is separated by $3.0\,$arcsec from a FIRST source and the
quasar lies along a line joining the catalogued FIRST source to a
second FIRST source, too faint to be included in the FIRST source
catalogue but clearly visible in the FIRST image cutout and almost
coincident with the quasar. All three of these LBQS quasars are
virtually certain to be associated with the multiple FIRST sources.
The predicted number of chance LBQS--FIRST coincidences within a
$10\,$arcsec radius is $\simeq 2$. Two further LBQS quasars (B1138-0126
and B2359-0216B) have single catalogued FIRST sources with separations
between $5$ and $10\,$arcsec and they most likely reflect chance
superpositions. No further discussion of these two objects is included
in the paper.

Table 2 includes details of the LBQS--FIRST source matches. Column 1,
LBQS name, derived from the B1950.0 coordinates; columns 2--3, the
J2000.0 right ascension and declination for the LBQS objects derived
from the UKST plates; column 4, redshift; column 5, $B_J$ magnitude;
column 6; FIRST source associated with the LBQS object; column 7,
integrated radio flux from the FIRST catalogue; column 8, associated
rms uncertainty in the FIRST maps at the source position; column 9,
angular separation between the LBQS and FIRST sources; column 10, ${\rm
log} R^*$ the $k$--corrected ratio of radio to optical flux for each
optical--radio association. For completeness, the three LBQS sources
with optical--radio separations $> 2.0\,$arcsec, denoted by ``N'' in
column 11, are listed in Table 2 but these objects are not
included in the assessment of the LBQS--FIRST match statistics.

The values of ${\rm log} R^*$, the $k$--corrected ratio of radio to
optical flux for each optical--radio association, employed herein
have been calculated according to the definition of Stocke et al.
(1992). To facilitate comparison with the results of the FIRST quasar
survey the same values for the assumed power--law spectral energy
distributions, $F(\nu) \propto \nu^{\alpha}$, in the radio and optical,
$\alpha_r =-0.5$ and $\alpha_o=-1.0$ respectively, have been adopted as
in White et al. (2000). In calculating ${\rm log} R^*$ White et al.
employ the photographic $O$ magnitude as a direct estimate of the
conventional $B$ magnitude employed in the definition. It is possible
to make a series of quite complicated corrections to obtain an estimate
of the quasar continuum flux at $\sim 4500\,$\AA \ from the $B_J$
magnitudes (Hooper et al. 1995; \S 2.3). However, over the redshift
range $0.2 \le z \le 3.4$ applicable to this investigation, the mean
$B-B_J$ for a typical quasar is $B-B_J \simeq 0.07$, which results in
an increase of only $0.03$ in ${\rm log} R^*$. Thus, differences in
${\rm log} R^*$ due to bandpass effects are in general very small and
the uncertainty in the individual values of ${\rm log} R^*$ are
dominated by possible intrinsic variability over the $\sim 15\,$year
epoch difference between the acquisition of the radio and optical
fluxes and the errors in the radio and optical fluxes. Given these
circumstances the raw $B_J$ magnitudes have been adopted as a direct
estimate of $B$ magnitude.

\subsection{Matching the LBQS input catalogue and the FIRST catalogue}

Based on the predicted accuracy of the celestial positions of both
catalogues, together with the empirical verification provided by the
cross--correlation of the known LBQS quasar positions with the FIRST
catalogue, a pairing radius of $2.1\,$arcsec was adopted as the
criterion for a match between an optical and FIRST source. There are
$\sim 25\,000$ FIRST sources in the $270\,$deg$^2$ of overlap with the
LBQS catalogue. The LBQS flux--limited input catalogue contains $\sim
100\,000$ stellar images and $\sim 40\,000$ non--stellar images in the
region of overlap and the predicted number of chance coincidences
between FIRST sources and LBQS stellar and non--stellar sources are
$\simeq 10$ and $\simeq 4$ respectively.

The cross--correlation of the two catalogues produces a total of $596$
matches within the $2.1\,$arcsec pairing radius. Of these, $101$ are
classified as stellar by the APM and $495$ non--stellar.  The nature of
the objective--prism plate material employed to compile the LBQS
catalogue is such that a fraction of the flux--limited object catalogue
defined from the APM scans of the UKST direct plates cannot be
classified reliably. For example, objects may possess spectra
contaminated by spectra of nearby sources. This straightforward loss of
objects is responsible for the reduced effective area of the survey
relative to the total area of sky surveyed. For the 11 fields included
in the LBQS--FIRST overlap the predicted percentage of stellar and
marginally resolved objects not included in the LBQS is $17\%$ (Paper
VI; Table 2). Of the $101$ stellar FIRST--LBQS matches, $18$ ($17.8\%$)
are classified as unprocessable, in excellent agreement with the
prediction from the effective area calculation. The fraction of low
surface brightness galaxies and significantly resolved galaxies with
unprocessable spectra is in fact somewhat higher, although this is not
relevant for the LBQS quasar catalogue with its low redshift limit of
$z=0.2$, and $115$ of the $495$ non--stellar objects ($23.2\%$) are
classified as unprocessable.

Thus, the cross--correlation of the two catalogues produces 83 stellar
sources and 380 non--stellar sources that were searched as part of the
LBQS survey and that pair to catalogued FIRST sources. Of the 83
stellar sources 64 possess identification spectra obtained as part of
the LBQS: 62 quasars and unknowns (Table 2), 1 active galactic nucleus
and 1 star. Just 19 stellar objects, of which $\simeq 10$ are expected
to be chance coincidences, remain for which identification spectra were
not obtained as part of the LBQS.

The large number of non--stellar matches appears to present a less
tractable problem but in fact the majority of the objects can be
eliminated as potential quasars from consideration of their image
profiles on the UKST plates. The LBQS survey was designed to be
sensitive to quasars with both unresolved and resolved images on the
UKST direct plates.  ``Resolved quasars'' might include objects at
relatively low redshift, where the host galaxy is also visible, as well
as gravitationally lensed quasars. The objective--prism spectra of
objects, whether resolved or unresolved, satisfying the flux limits on
the UKST direct plates were searched as part of the LBQS survey.
However, the very tight, monotonic, relation between the peak and
integrated brightness of unresolved stellar images on the UKST direct
plates means it is possible to define a simple criterion that
eliminates a large fraction of the resolved images from consideration.
Specifically, for a resolved object to contain an unresolved source
bright enough to satisfy the LBQS flux limits, the image of the resolved
object must contain pixels that achieve a peak brightness at least
equivalent to that of an unresolved source at the faint limit of the
LBQS flux--limited catalogue (Paper VI: \S3.1.3). The non--stellar
sources may thus be divided into two categories, i) images with a peak
surface brightness too faint to include an unresolved source bright
enough for inclusion in the LBQS, 222 objects, and ii) images whose
peak surface brightness exceeds that of the faintest unresolved source
in the LBQS flux--limited catalogue, 158 objects. Five of the objects
in the latter category had been identified as quasars (Table 2) and 4
as active--galactic--nuclei during the LBQS, leaving 149 objects for
which identification spectra were not obtained as part of the LBQS.

Given the small predicted number of chance coincidences the majority of
the non--stellar optical--radio associations likely reflect physical
association. However, the peak surface brightness criterion used in the
definition of the sample is extremely conservative. The vast majority
of the ``high surface brightness'' non--stellar objects possess
extended morphologies, consistent with galaxies at low redshift, $z
\sim 0.1$, and no significant unresolved component exists. A
significant fraction of the small number of objects with redshifts that
exceed the LBQS low--redshift limit of $z=0.2$ and that show evidence
for active--galactic nuclei, will also not possess unresolved
components bright enough to be included in the LBQS. The nature of the
small number of such non--stellar objects is further discussed in \S 4, following a description of the spectroscopic observations of the sources.

\section{SPECTROSCOPIC OBSERVATIONS}

Spectroscopic observations of 11 stellar and 35 non--stellar candidates
were obtained using LRIS, the Low Resolution Imaging Spectrograph (Oke
et al.  1995), on Keck I during the night of 2000 April 30. A 300
line/millimetre grating, blazed to $5000\,$\AA, and a slit--width of
$1\,$arcsec gave a wavelength coverage of $\lambda\lambda 4550-9500$ at
a resolution of $7\,$\AA. Standard bias and flatfield exposures were
obtained at the start of the night and comparison lamp exposures were
obtained every few hours to monitor the wavelength stability of the
spectrograph.  Wavelength stability was determined to be $\la
2.5\,$\AA.  The slit was aligned East--West throughout the
observations. Exposures of three spectrophotometric standard stars were
obtained in order to place the spectra of the candidate objects on a
relative flux scale but no attempt was made to obtain absolute
spectrophotometry. Transparancy, which was good, and the seeing of
$\sim 1\,$arcsec were stable throughout the observations.  Exposure
times of $300\,$s for the stellar objects predicted to be quasars and
$120-180\,$s for the other objects produced spectra of excellent
quality, S/N$\sim 20$, and the identifications, based on multiple
absorption and emission features, are unambiguous.  Redshifts for the
extragalactic sources were determined by cross-correlation of the
quasar spectra with the LBQS composite spectrum (Francis et al, 1991)
and by direct measurement of the wavelengths of narrow emission lines,
when present, or absorption features for the galaxies.

R. Becker and collaborators provided the spectrum of one stellar candidate
identified as a quasar during their follow--up program of FIRST
sources. In 1999 November they very generously also undertook LRIS
observations on Keck I, employing a set--up very similar to that
described above, of 6 of the stellar candidates, confirming their
identifications as stars (Becker 1999, private communication).

Further spectroscopic observations of 1 stellar and 26 non--stellar
candidates were obtained with the 6.5m MMT on the nights of 2000
November 19--20.  The Blue Channel of the MMT Spectrograph was used
with a 500 line/millimetre grating, blazed at $5410\,$\AA, and a
slit--width of $1.5\,$arcsec, giving a wavelength coverage of
$\lambda\lambda 4100-7500$ at a resolution of $\sim5\,$\AA.  The slit
was aligned along a constant azimuth to minimize the effect of
atmospheric dispersion. Calibration, data reduction, and measurement of
the redshifts followed procedures essentially identical to those for
the Keck data.

The 2dF Galaxy Redshift Survey (2dFGRS) (Colless 1999) has a
considerable overlap with the 6 LBQS fields that lie along the
equatorial strip in the North Galactic Cap. The 2dFGRS Team have
kindly made available the redshifts of 31 non--stellar objects listed
in their catalogue as of 2000 December.  Finally, a ``Near Position''
search of the NASA Extragalactic Database produced 7 additional
galaxy identifications with published redshifts.

Combining the spectroscopic identifications from our own observations,
the 2dF Galaxy Redshift Survey, R. Becker (and collaborators) and the 
NED--search produces identifications for all the stellar sources
and for more than two--thirds of the non--stellar sources.

Table 3 contains the positions, radio properties and spectroscopic
classifications of the 21 stellar objects in the LBQS flux--limited
sample that are not included (as LBQS quasars or BL Lac candidates) in
Table 2. Columns 1--2 , the J2000.0 right ascension and declination for
the optical source derived from the UKST plates; column 3, $B_J$
magnitude; column 4; FIRST source associated with the LBQS object;
column 5, integrated radio flux from the FIRST catalogue; column 6,
associated rms uncertainty in the FIRST maps at the source position;
column 7, angular separation between the LBQS and FIRST sources;
column 8, LBQS classification (if available) of the source prior to
observation; column 9, spectroscopic identification; column 10,
redshift; column 11, source of the spectroscopic classification.

Table 4 gives the positions and radio properties for all 153
non--stellar objects in the LBQS flux--limited sample that are not
included (as LBQS quasars or BL Lac candidates) in Table 2.
Spectroscopic identifications are included for 103 objects. Columns
1--2, the J2000.0 right ascension and declination for the optical
source derived from the UKST plates; column 3, $B_J$ magnitude; column
4; FIRST source associated with the LBQS object; column 5, integrated
radio flux from the FIRST catalogue; column 6, associated rms
uncertainty in the FIRST maps at the source position; column 7,
angular separation between the LBQS and FIRST sources; column 8,
galaxy redshift or spectroscopic classification; column 9, source of
the spectroscopic classification. The references for the redshifts of
the 7 galaxies identified in the NED--search are indicated.

The information for each non--stellar candidate is identical to that
given for the stellar candidates in Table 3 with the following
exceptions. None of the non--stellar sources were identified as LBQS
quasar candidates and the pre--identification LBQS classification is
therefore omitted. The procedure used to generate a close to linear
relation between APM instrumental magnitudes and photometric magnitudes
is based on the universal profile of the unresolved sources on each
UKST plate (Bunclark and Irwin 1983). The scheme is extremely effective
for the stellar sources but for resolved objects, whose surface
brightness profiles exhibit a very wide range of shape, the derived
``photometric'' magnitudes possess a large scatter and are
systematically too bright, often by as much as a magnitude or more.
Thus, while all the objects listed in Table 4 possess instrumental
magnitudes that place them within the LBQS flux--limited sample, only
in the cases where the image morphology is close to stellar will the
photometric magnitudes be reliable. The photometric magnitudes have
been included in Table 4 for completeness but for non--stellar objects
the values are bracketed and specified to only one decimal place. A
discussion of those objects with redshifts exceeding the LBQS low
redshift limit of $z=0.2$ is included in \S4.2.

\section{NEW LBQS--FIRST QUASARS}

\subsection{The stellar LBQS--FIRST objects}

The identification of 10 objects, most with optical--radio separations
$\ge 1\,$arcsec, as Galactic stars accords well with the predicted
incidence of chance associations (\S2.4). Seven objects have redshifts
$z> 0.2$, placing them above the lower redshift limit of the LBQS, and
1 object has a featureless spectrum consistent with an identification
as a BL Lac. The status of two of these objects deserves some comment.

We note with some embarrassment that we acquired a spectrum of
J1022--0103 at the MMT in 1990 March, early in the compilation of the
LBQS catalogue. On the basis of that spectrum the object was classified
as a ``star''.  While considerable attention was paid to ensuring that
``star'' classifications were assigned only to objects with positive
identifications of stellar features, J1022--0113 was mistakenly so
classified and it should have appeared in Paper VI as an ``unknown''.
The very close optical--radio match and large radio--flux indicate the
object is almost certainly a BL Lac. Indeed, identification with the
bright ROSAT source RXJ1022.7-0112 (Voges et al. 1999) led Appenzeller
et al. (1998) to identify J1022--0103 as a BL Lac.

The spectrum of J1438+0032 ($z=0.209$), Figure 2, shows a well--defined
continuum, of essentially neutral colour, with stellar absorption
features, including Ca H+K, G--band and Na D, visible.  Strong, narrow
and asymmetric emission lines, including H$\beta$, [\ion{O}{3}]
$\lambda\lambda$4959,5007, H$\alpha$+\ion{N}{2} $\lambda$6584, [\ion{S}{2}]
$\lambda\lambda$6716,6731 and the absence of any broad line component
are consistent with the presence of a Seyfert 2 nucleus.
Notwithstanding the APM--classification of ``stellar'', the object on
UKST plates is clearly a galaxy with a compact appearance that does not
appear to possess an unresolved source bright enough to satisfy the
LBQS flux limit. The integrated magnitude $B_J=18.33$ is close to the
LBQS flux limit of $B_J=18.41$ in the field and the magnitude of $B_J
\ge 18.47$, based on the peak brightness, places the nuclear component
below the LBQS flux limit. The presence of detectable stellar
absorption features, in what is essentially a nuclear spectrum
(acquired with a $1\,$arcsec slit), is consistent with the conclusion
that a substantial fraction of the integrated light derives from
stars.  This conclusion is further supported by the form of the
objective--prism spectrum, which extends much bluer ($\sim 3500$\AA)
than the $\sim 4600$\AA \ limit of the Keck spectrum. The
objective--prism spectrum is dominated by a well--defined component in
the wavelength range $4500-5300$\AA, indicative of a $z\simeq 0.2$
early--type spiral. A definite ultraviolet tail extends to the
atmospheric cutoff but it contains only a small fraction of the total
flux. We conclude that the active galactic nucleus in J1438+0032 falls
significantly below the flux--limit for inclusion in the LBQS. Thus,
the complete spectroscopic identification of all ``stellar'' sources
matched to the FIRST catalogue produces 6 new quasars and one
(previously misclassified) BL Lac object.

\subsection{The non--stellar LBQS--FIRST objects}

Five of the 103 non--stellar objects with spectroscopic observations
have redshifts $z > 0.2$, all just above the lower redshift limit of
the LBQS. To satisfy the criteria for inclusion in the LBQS an object
must also contain an unresolved nuclear component brighter than the
faint LBQS $B_J$ flux limit. The morphologies and spectra of three
galaxies confirm that any nuclear component is much fainter than the
LBQS flux limit in the relevant UKST field: J1027--0216 ($z=0.218$)
possesses a spectrum with significant stellar absorption features (Ca
H+K, G--band, Na D...). The presence of strong, narrow, emission lines
of [\ion{O}{3}] $\lambda\lambda$4959,5007 and [\ion{O}{2}]
$\lambda$3727, together with weak H$\beta$ emission indicate the object
is experiencing a period of star--formation. The image morphology is
extended and any nuclear component must be significantly fainter than
the LBQS flux limit of $B_J=18.68$ in the field.  The 2dFGRS magnitude
is $B_J=18.86$, confirming the lack of any bright component. J1334+0102
($z=0.203$) also possesses a spectrum with significant stellar
absorption features and H$\gamma$ and H$\delta$ in absorption. Evidence
for a relatively recent episode of star formation comes from the
presence of weak [\ion{O}{3}] $\lambda\lambda$4959,5007 and H$\beta$
emission along with stronger [\ion{O}{2}] $\lambda$3727 emission. The
overall form of the spectrum is consistent with that of an early--type
spiral galaxy.  The image morphology is significantly extended and
there is no evidence for a nuclear component bright enough to satisfy
the LBQS flux limit of $B_J=18.41$ in the field. The 2dFGRS magnitude
of the object is $B_J=18.83$.  A brighter galaxy is located $\sim
15\,$arcsec to the south--east. J1449--0116 ($z=0.202$) has a spectrum
completely dominated by an old stellar population. There is no evidence
for emission lines or a non--stellar continuum. The morphology of the
object is entirely consistent with that of a $z=0.2$ early--type galaxy
and there is no evidence for the presence of an unresolved nuclear
component brighter than the LBQS flux limit of $B_J=18.62$ in the
field. The 2dFGRS magnitude is $B_J=18.38$.

The two other non--stellar objects with $z > 0.2$, J0100--0200
($z=0.227$) and J1334+0102 ($z=0.246$) are very different in character.
Both possess strong blue continua, broad hydrogen lines, weaker narrow
oxygen lines and stellar absorption features are not evident. The MMT
spectrum of J0100-0200 extends far enough to the blue that strong broad
Mg II $\lambda2798$ emission can just be seen. While the objects are by
definition non--stellar, their image morphologies are compact on the
UKST plates and the presence of bright unresolved components was noted
during their acquisition at the telescopes.  Magnitudes, estimated from
the peak brightness in the images, for both objects are constrained to
be $B_J \ge 18.2$. The nuclear component will be several tenths fainter
than this bright limit, however, J0100--0200 will certainly satisfy the
associated LBQS flux limit of $B_J=18.77$. The magnitude of J1334+0102
will place it very close to the LBQS flux limit of $B_J=18.41$ in its
field but we shall assume that it does indeed qualify for inclusion in
the LBQS. Thus, the spectroscopic identification of two--thirds of all
the ``non--stellar'' sources matched to the FIRST catalogue produces 2
new quasars.

The optical and radio properties of the total of 9 new objects
satisfying the LBQS redshift and flux selection criteria are listed in
Table 5.  Column 1, name, derived from the J2000.0 coordinates; columns
2-3, the J2000.0 right ascension and declination from the UKST plates;
column 4, redshift; column 5, $B_J$ magnitude; column 6; FIRST source
associated with the LBQS object; column 7, integrated radio flux from
the FIRST catalogue; column 8, associated rms uncertainty in the FIRST
maps at the source position; column 9, angular separation between the
LBQS and FIRST sources; column 10, ${\rm log} R^*$ the $k$--corrected
ratio of radio to optical flux for each optical--radio association;
column 11, LBQS classification (if available) of the source prior to
observation; column 12, spectroscopic identification. The LBQS
candidates were classified on a scheme indicating decreasing
probability an object was a quasar, specifically: ``QSO'', ``QSO?'',
``QSO??'', ``Star??'', ``Star?'', ``Star''. Thus 6 of the 9 new objects
were identified as quasar candidates during the compilation of
the LBQS survey. The optical spectra of all 9 objects are shown in
Figure 3.

The spectra of the 9 new objects are largely unremarkable, with
continuum and emission line properties very similar to a number of
existing LBQS quasars. However, 2 objects possess absorption features
of some note: J1316--0038: shows two strong and moderately broad
absorption troughs near the peak of the Mg II emission line.  The
presence of this absorption affects the redshift derived from
cross--correlation, $z_{em} =0.895$.  A narrow emission feature,  seen
at at $\lambda7192.5$, could be due to [\ion{O}{2}] $\lambda3727$.
Assuming this identification is correct, the systemic redshift of the
quasar is $z=0.930$ and the longer wavelength absorption trough has
$z_{abs} > z_{em}$. J1333+0012: two narrow Mg II absorption doublets
are seen within $15\,000\,$km s$^{-1}$ blueward of the Mg II emission
line peak.

\section{EFFECTIVENESS OF THE LBQS}

The discussion concerning the effectiveness of the LBQS survey given in
Paper VI concluded that the survey contained $\simeq 90\%$ of the total
number of quasars satisfying the redshift and flux--limits of the
survey. The majority of the quasars not included in the published
survey were believed to lie among the lower--probability candidates
(objects classified Star??, QSO?? or QSO?) and for which spectroscopic
observations had yet to be made. Attention was also drawn to the
redshift interval $z = 0.6 - 1.0$, in which the properties of typical
quasar SEDs produce objective--prism spectra that are relatively red
and largely featureless and where the survey was therefore predicted to
be less effective than average. Consideration of the nature of the
candidate selection algorithms and empirical comparison with surveys
and compilations of existing quasars led to the conclusion that only
very red, essentially featureless objects, such as red BL Lacs, would
escape detection.

The direct comparison of the entire LBQS flux--limited source catalogue
with the FIRST survey over more than half the LBQS survey area,
summarised in Table 6, provides a stringent test of the optical survey
and the conclusions made concerning its effectiveness. Based on the
confirmation of 2 quasars among the 103 non--stellar objects with
spectroscopic identifications we assume that one further quasar will be
confirmed among the 50 remaining non--stellar objects without
spectroscopic observations listed in Table 4.  Based on this
assumption, the total number of FIRST--detected quasars (and BL Lacs)
in the LBQS flux--limited catalogue is 77 (67 objects from Table 2, 9
objects from Table 5 and the 1 object inferred to lie among the
non--stellar objects without spectroscopic redshifts in Table 4).  The
straightforward fractional incompleteness is thus $13\pm4\%$, in good
agreement with the prediction from Paper VI.

The radio properties of the new objects can be compared to those of the
67 LBQS--FIRST objects previously known. Figure 4 shows ${\rm log} R^*$
as a function of redshift for the objects from Table 2 and Table 5. A
Student t--test and Kolmogorov--Smirnov test give probabilities of 0.20
and 0.12, respectively that the samples are drawn from the same
distribution. The similarity of the apparent magnitude distributions
also supports the conclusion that there is no significant difference in
the gross properties of the two samples: A Student t--test and
Kolmogorov--Smirnov test for the distribution of apparent magnitudes
give probabilities of 0.52 and 0.63 respectively that the samples are
drawn from the same distribution.

The consistency of the distributions of ${\rm log} R^*$ and apparent
magnitude is not surprising given the status of the 9 new LBQS--FIRST
objects within the LBQS candidate list. Six of the 9 objects appear in
the LBQS candidate list, indeed one of the objects (J1022-0113) was
observed spectroscopically. J1143+0113 ($z=1.282$), for example, is an
extremely blue object that was identified at an early stage of the LBQS
but the noise in the only moderate signal--to--noise ratio
objective--prism spectrum conspires to produce apparent absorption
lines at wavelengths corresponding to H$\beta$ and H$\delta$, suggesting
classification as a hot white dwarf, and the object was therefore not
assigned a high priority for spectroscopic follow--up. Figure 5
illustrates the colours of the objects from both Tables 2 and 5, along
with colours for all the remaining LBQS quasars, as a function of
redshift. The definition of the colour--measure, the Half--Power Point
(HPP) of the objective--prism spectra, is described in \S 3.1.1 of
Paper VI. In this representation, the dashed--line at HPP=0 indicates
the boundary below which objects satisfy the basic colour selection
criterion for inclusion in the LBQS candidate list. The substantial
spread in colour at fixed redshift is well illustrated by the
distribution of the LBQS quasars as a whole. The systematic variation
in the location of the mid--point of the distribution as a function of
redshift is directly analogous to the well--established variation in
$U-B$ colour as a function of redshift exhibited by quasars. The extent
of the spread in colour does not vary significantly as a function of
redshift, although a small fraction of pathological objects does exist.
The conclusion drawn in Paper VI regarding the likely reduction of the
effectiveness of the LBQS selection in the redshift region about
$z=0.8$ was based on noting that, assuming an essentially constant
dispersion in HPP as a function of redshift, a number of the redder
quasars at $z\sim 0.8$ would fail to satisfy the colour selection and
in many cases their spectra do not possess other characteristic
features that would allow their identification as high--probability
candidates. The confirmation that 4 of the 9 new LBQS--FIRST objects
have redshifts in the range $z=0.7-1.0$ thus accords well with
expectations.

The other redshift interval where increased difficulty in identifying
quasars is not unexpected is close to the low--redshift limit, $z=0.2$,
of the survey.  The low--redshift cutoff of $z=0.2$ for the LBQS was
chosen with some care.  While the candidate selection algorithms are
applied to objects irrespective of morphology, the presence of a
significant contribution of stellar light from the host galaxy causes
the fainter quasars at $z \la 0.2$ to appear increasingly red. A
discussion of the problem in the context of identifying quasars in
bright host galaxies, including the results of simulations of
quasar+host--galaxy spectra, is contained in Hooper et al. (1995;
Appendix B). Figure 9 of Hooper et al. illustrates how rapidly the
colour of quasar+host systems moves to the red at $z \sim 0.2$. The
restriction of the LBQS survey selection function to nuclear components
that satisfy the faint flux limit of the LBQS, combined with the
continuity in the amplitude of the spread of HPP as a function of
redshift seen in Figure 5, means that it is extremely unlikely that a
significant population of quasars at $z\ga 0.2$ have been excluded from
the LBQS. However, the number of objects in the flux--limited source
catalogue with HPP values close to the selection boundary is large and
the detection of 2 new LBQS--FIRST objects with $z<0.25$ likely
reflects a genuine decrease in the effectiveness of the selection right
at the redshift limit. As is evident from Figure 5, any such effect
is confined to $z\la 0.25$ as the locus of the HPP distribution moves
rapidly blueward, dropping below the HPP=0 line boundary, even by
redshift $z=0.3$.

In summary, the small number of additional LBQS--FIRST objects possess
gross optical and radio properties entirely consistent with those of the quasars
previously identified in the LBQS. The number and redshift distribution
of the additional LBQS--FIRST objects also accords well with the
predictions of the effectiveness of the LBQS selection criteria. The
conclusion that no new population of objects with significantly
different properties has eluded the LBQS optical selection is borne out
by consideration of the spectra of the 9 LBQS--FIRST objects shown in
Figure 3. In the context of the LBQS survey the spectra are
unremarkable both individually and taken together as a sample. For each
object, examples of quasars with essentially identical SEDs at
comparable redshifts exist within the LBQS survey.

\section{DISCUSSION}

Notwithstanding advances in the application of quantitative techniques
to the selection of quasars at optical wavelengths, suggestions that
optical surveys suffer from significant biases against the detection of
certain types of quasar persist.  For example, in \S5.3 of their
presentation of the FIRST Bright Quasar Survey (FBQS), White et al.
(2000), discuss such a bias in the context of the detection of
low--ionization BAL QSOs. The surface density of the recently--detected
``radio--loud'' BALs among the FBQS quasars is too low for the
LBQS--FIRST match presented in this paper to provide interesting
constraints. However, the essentially complete census of 76
radio--detected objects within the LBQS flux limits presented here
demonstrates conclusively that there is no significant bias against any
class of quasar, constituting $\ga 7\%$ of the FIRST radio--detected
population\footnote{The limit of $7\%$ is derived by noting that the
detection of no such objects constrains the true incidence of any
such population in the survey area to be $\la 5$ objects}.  The same
constraint applies to any hypothetical bias at particular redshifts or
magnitudes, other than the previously reported reduction in the
efficiency of the LBQS at $z\sim 0.8$. Furthermore, the FIRST--LBQS
comparison offers no support for Usher \& Mitchell's (2000) contention
that the LBQS is unsuitable for the estimate of the quasar luminosity
function at low redshifts. Sample limits of redshift $z \ge 0.2$ and
magnitude $B_J \ge 16.5$ should be adopted for such statistical
analyses (Hewett, Foltz \& Chaffee 1993, Paper VI). To facilitate
quantitative comparison between the properties of the LBQS catalogue
and other surveys we have adopted a completeness as a function of
redshift as tabulated in Table 7. The numerical values are based on
adopting a global incompleteness of $10\%$ with an increase in the loss
of objects at redshifts $z=0.75-0.85$ and at the low redshift limit,
$z=0.2$, of a factor of two.

Perhaps most importantly, the results of the LBQS--FIRST comparison
offers the prospect of utilising both the LBQS and FBQS, with their
very large numbers of quasars, to better understand the physical origin
of the very broad range of SEDs within the quasar population. The
analysis of composite spectra from the two surveys by Brotherton et al.
(2001) signals the start of this process. However, the optical
bandpasses employed in the selection of the two samples, $B_J$
($\lambda_{eff} \simeq 4600$\AA, $\Delta\lambda \simeq 1100$\AA) for
the LBQS and Palomar $E$ ($\lambda_{eff} \simeq 6400$\AA,
$\Delta\lambda \simeq 400$\AA) for the FBQS, are significantly
different and a careful treatment of the effects of the different
bandpasses used to define the optical flux limits is required when
interpreting the statistical properties of the quasar samples.

The potential importance of the passband differences is evident from
the very different fraction of radio--detected objects found to the
respective optical flux limits in the two surveys.  White et al. (2000)
suggest that at the brightest optical magnitudes the surface density of
optically--detected and radio--detected quasars are essentially
identical.  Even at the faint optical limit of the FBQS, $E=17.8$,
White et al. conclude that the radio--detected fraction is $\simeq
25\%$.  The FIRST--LBQS comparison in this paper provides a hard upper
limit to the fraction of the quasar population with optical magnitudes
$16.0 \le B_J \la 18.7$ detected in the FIRST survey. Based on the
total of 570 previously--known LBQS quasars and ``unknowns'' within the
FIRST survey area and the statistics of the LBQS--FIRST match (67
radio--detected objects among the 570 LBQS quasars and ``unknowns'' and
10 new radio--detected objects), the fraction of quasars detected by
FIRST is $13\%$ (i.e., 77/(570+10)). An improved estimate is obtained
by applying an incompleteness correction to allow for the fraction of
quasars not included in the LBQS. Adopting a survey completeness of
$90\%$ for the number of LBQS quasars which are not detected in the
radio (i.e., 503 objects), reduces the percentage of radio--detected
objects to $12\pm2\%$, very different from the figure of $25\%$ of
objects with magnitudes $E \le 17.8$ deduced by White et al. (2000).
Dividing the LBQS into magnitude bins containing $\sim 100$ quasars,
Table 8, shows a trend in the same sense as that found by White et al.
but the fraction of radio--detected objects\footnote{The true
fractions, allowing for the $\simeq 10\%$ incompleteness in the census
of LBQS quasars only detected in the optical, will decrease by $\sim
0.01$} is significantly smaller than they find at all magnitudes. For
the very brightest LBQS objects, $16 \le B_J \le 17.0$, where the
statistics are poor (14 quasars, 6 radio--detected, fraction 0.43), the
radio--detected surface density approaches half the optical surface
density. So we conclude that only for the optically very brightest
quasars, $B \la 16$, can White et al.'s claim that the radio--detected
surface density matches that of optically--selected quasars be true.

The explanation for the large differences in the radio--detected
fraction of quasars may be due to a combination of two
factors: (i) radio--detected quasars possess somewhat redder SEDs
compared to the population as a whole and the $E$--magnitude limit
therefore results in a larger fraction of radio--detected objects
compared to the LBQS sample limited by the much bluer $B_J$ passband,
(ii) the form of the conversion employed by White et al. to relate
published quasar surface densities from flux--limited samples at blue
wavelengths to the FBQS surface--densities determined in the
$E$--passband. Employing a large database of broadband colours, $B$
though $K$, we intend to present a detailed investigation of
differences in the SEDs of sub--samples of quasars from the LBQS and
FBQS in a future paper.  However, preliminary indications of the origin
of the difference in radio--detected fractions can be drawn using
composite spectra for the two samples.  We have used the FBQS composite
spectrum derived from 657 FBQS quasars and the LBQS composite employed
by Brotherton et al.  (2000).  Observed--frame spectra over the
redshift range $0.2 \le z \le 3.0$ were generated using both
composites. These were then transformed to simulate their appearance on
the objective--prism plates employed in the LBQS survey.  The procedure
is based on an empirical transformation derived from $\sim 2000$
flux--calibrated quasar and stellar spectra and their associated
spectra on the objective--prism plates.  The HPP values for the two
composite spectra are shown in Figure 6 (solid line \---\ LBQS
composite; dashed line \---\ FBQS composite).

The extremely good match between the centroid of the distribution of
HPP values for the individual LBQS quasars and the synthetic locus of
the LBQS composite is evident. The only significant difference between
the loci of the LBQS and FBQS composites occurs in the redshift
interval $1.8 \la z \la 2.2$ where the stronger Ly$\alpha$ emission
discussed by Brotherton et al. causes the FBQS composite to become
bluer than the LBQS composite. The FBQS composite is marginally redder
than the LBQS composite at redshifts $z \la 1.3$ but the difference is
small relative to the intrinsic spread among the individual quasars and
relative to the distance of the loci from the LBQS colour--selection
boundary. There is no evidence for any significant difference in the
distribution of colours of the 76 radio--detected quasars compared to
the LBQS quasars as a whole. If the SEDs of the radio--detected objects
were significantly redder (over the observed--frame wavelength interval
$3500-5300$\AA) the objects would possess larger HPP values at a given
redshift. The similarity of the LBQS and FBQS composite spectra,
coupled with the very similar distributions of quasars with and without
radio detections in the LBQS colour--selection space, suggests that
systematic differences between the SEDs of radio--detected quasars and
the population as a whole are not large.

The availability of the FBQS and LBQS composite spectra allows a more
direct estimate of the surface density of radio--detected quasars for a
sample flux--limited in the Palomar $E$ passband. Table 9 lists the
$B_J-E$ and $O-E$ colours for the FBQS and LBQS composite spectra as a
function of redshift. The maximum redshifts are set by the point at
which the Lyman--$\alpha$ forest begins to affect the flux in the $B_J$
and $O$ passbands significantly. The composite spectra are, by their
nature, made up from quasars with an extended range of redshift and the
depression of the continuum at rest--frame wavelengths $\la 1216$\AA
\ does not reproduce the behaviour seen in actual quasars at the higher
redshifts.  The colours were calculated using the Palomar $O$ and $E$
sensitivity curves of Minkowski \& Abell (1963) and our own
determination of the $B_J$ sensitivity. The filter plus emulsion
sensitivity curves were combined with one reflection off aluminium and
atmospheric absorption and extinction appropriate for observations made
at an airmass of 1.3 for a relatively low--altitude site such as Siding
Spring or Mount Palomar. The synthetic photometry was performed using
the {\tt synphot} package in the Space Telescope Science Data Analysis
System.

The $B_J-E$ colours show significant variations with redshift,
including a prominent redward excursion at $z \simeq 0.3$ where
[\ion{O}{3}] $\lambda\lambda$4959,5007 and H$\beta$ emission moves
through the narrow $E$--passband. Additional redshift increments have
been included in Table 9 to quantify the rapid change in the colours at
$z\simeq 0.3$. The $\sim 0.25$ magnitude reddening in the $B_J-E$
colours coincides exactly with the excess of quasars seen at this
redshift in the FBQS redshift distribution (White et al. 2000;
Figure 12).

Taking a redshift range of $0.2 \le z \le 3.0$, with the lower limit
set by the LBQS low--redshift boundary and the upper limit taken so
that the FBQS colour--selection criterion of $O-E \le 2.0$ is not a
significant factor, we estimate the true surface density of FBQS
quasars at $15.5 \le E \le 17.8$ as follows. The 565 quasars in Table 2
of White et al., within an area of $2682\,$deg$^2$ give a surface
density of $0.211\,$deg$^{-2}$. This number is reduced by a factor of
1.09 to take account of the Eddington--bias (Eddington 1913) due to an
rms magnitude error of $\sigma_E=0.2$, giving a final surface density
estimate of $0.193\,$deg$^{-2}$.

For the LBQS, we take the quasars in the extended sample, calculate the
$E$ magnitudes using the original $B_J$ magnitudes and the $B_J-E$
colours for the FBQS composite in Table 9, weighting each quasar by the
reciprocal of the completeness estimate as a function of redshift
(Table 7). the number of quasars with magnitudes $15.5 \le E \le 17.8$
is then calculated.  A reduction of a factor 1.02 is made to account
for Eddington--bias due to an rms magnitude error of
$\sigma_{B_J}=0.1$.  The calculation was repeated using the $B_J-E$
colours for the LBQS composite spectrum. The resulting surface density
estimates are $1.252\,$deg$^{-2}$ (FBQS composite colours) and
$1.145\,$deg$^{-2}$ (LBQS composite colours), giving a radio--detected
fraction of $15-17\%$.

The calculation is far from ideal as no account is taken of the
dispersion of SEDs among the quasars, i.e., all quasars are assumed to
possess the same SED. However, the insensitivity of the surface density
estimate to which of the two composite spectra are used to calculate
the $B_J-E$ colours is encouraging and the resulting radio--detected
fraction is much closer to the direct estimate of $12\%$ found in this
paper (c.f., the fraction of $25\%$ advocated by White et al.). 

The importance of establishing the transformation between magnitude
bands for studies of the luminosity function and evolution of the
quasar populations is evident from consideration of the redshift
distributions of the LBQS and FBQS samples. Figure 6a shows the
redshift histogram of the 565 FBQS quasars with redshifts $0.2 \le z
\le 3.0$ and magnitudes $15.5 \le E \le 17.8$ from Table 2 of White et
al.  (2000). The redshift histogram of the 1067 quasars in the extended
LBQS sample, after application of the redshift completeness factors
listed in Table 7, normalised to the same number of quasars as in the
FBQS is also shown.  Compared to the LBQS redshift distribution the
FBQS exhibits a significant excess of objects in the $z=0.2-0.4$ bin
and a much steeper decline in the number of quasars over the range
$z=1.4-2.0$. Brotherton et al.  (2001) suggest that the larger number
of quasars at low redshift in the FBQS may be due to the inclusion of
very bright sources in the FBQS.  In fact, whether the small number
(13) of very bright ($E < 15.5$) FBQS--sources are included makes no
significant change to the differences evident between the two
histograms. Figure 6b reproduces the FBQS redshift histogram together
with error bars based on the number of objects per bin. Also shown is
the redshift histogram for the sub--sample of LBQS quasars with $15.5
\le E \le 17.8$, where the $E$--magnitudes have been calculated
according to the procedure described above, together with associated
error bars. The LBQS histogram shown in Figure 6b was derived using the
$B_J-E$ colours for the FBQS composite spectrum but there is no
discernible difference if colours from the LBQS composite are employed
instead. Note the increased frequency of low--redshift, $z=0.2-0.4$,
quasars and the much steeper decline in the number of quasars at $z >
1.4$ in the transformed LBQS--histogram. Indeed, the form of the
redshift histograms over the full redshift range, $z=0.2-3.0$, are very
similar and we conclude that comparison of the distribution of quasars
in the FBQS and LBQS as a function of redshift adds further support to
the conclusions regarding the effectiveness of the LBQS and modern
optically--selected quasar surveys in general.

\section{CONCLUSIONS}

The principal results of this study can be summarised as follows:

1. Cross--correlation of the source catalogues for the LBQS and the
First Survey in $270\,{\rm deg}^2$ of sky common to both surveys
yields 67 positional coincidences between FIRST sources and known LBQS
quasars.  In addition, 19 stellar and 149 non--stellar coincidences are
found.  Optical spectroscopy of all of the stellar and two--thirds of
the non--stellar sources reveals eight new quasars and one BL Lac
object that was observed but misclassified during the LBQS.

2. The fractional incompleteness of the LBQS determined from this
sample is $13\pm 4\%$, in good agreement with the published estimate of
$10\%$.  Furthermore, four of the new objects have redshifts $0.7 \le z
\le 1.0$ where the LBQS selection techniques are known to have
decreased efficiency.

3. The distributions of radio--to--optical power, apparent magnitude
and spectroscopic properties of the new objects are consistent with the
67 previously--known LBQS--FIRST objects.  This essentially complete
census of 76 radio--selected objects within the LBQS flux limits
demonstrates conclusively that there is no significant bias in the LBQS
selection against any class of quasar constituting more than $\simeq 7\%$ of
the FIRST radio--detected population.

4. The fraction of radio--detected quasars in the LBQS catalogue is
found to be $12 \pm 2\%$, considerably smaller than the value of $25\%$
found by White et al. (2000) for the FBQS. To facilitate comparison of
the two surveys we present $B_J-E$ colours as a function of redshift
for the published LBQS and FBQS composite spectra.  Applying
corrections for the bandpass differences between the LBQS and FBQS
surveys we predict a radio--detected fraction of $15-17\%$ for the FBQS
sample, which is limited at $E=17.8$, close to the direct estimate of
$12\%$ found for the LBQS sample in this paper.

5. After correction for passband--dependent effects, the redshift
histogram over the redshift range $z=0.2-3.0$ for the sub--sample of
LBQS quasars with $15.5 \le E \le 17.8$ is very similar to that of the
FBQS.  This adds further support to our conclusions regarding the
effectiveness of the LBQS and modern optically--selected quasar surveys
in general.

We are particularly grateful to Bob Becker and collaborators for
undertaking spectroscopic observations on our behalf and allowing us to
reproduce their spectrum of J0103+0037.  The generosity of Matthew
Colless and the 2dF Galaxy Redshift Survey Team in providing galaxy
redshifts prior to publication and allowing us to include their
redshifts in this paper is greatly appreciated. This research has made
use of the NASA/IPAC Extragalactic Database (NED) which is operated by
the Jet Propulsion Laboratory, California Institute of Technology,
under contract with the National Aeronautics and Space Administration.
We are pleased to acknowledge the continued support provided for the
LBQS through NSF grant AST 98--03072. Data and analysis facilities at
the Institute of Astronomy were provided by the Starlink Project which
is run by CCLRC on behalf of PPARC. The authors wish to extend special
thanks to those of Hawaiian ancestry on whose sacred mountain we are
privileged to be guests.  Without their generous hospitality, many of
the observations presented herein would not have been possible.

\clearpage

\begin{deluxetable}{lcccccccc}
\tabletypesize{\footnotesize}
\tablewidth{0pt}
\tablenum{1}
\tablecaption{New LBQS Quasars}

\tablehead{
\colhead{LBQS Name}  & 
\colhead{R.A.} & 
\colhead{Dec.} & 
\colhead{R.A.} &
\colhead{Dec.} & 
\colhead{$z$} &
\colhead{$B_J$} &
\colhead{Field} & 
\colhead{Notes}\\
\colhead{} & 
\colhead{(B1950.0)} & 
\colhead{(B1950.0)} & 
\colhead{(J2000.0)} & 
\colhead{(J2000.0)} &
\colhead{} & 
\colhead{} & 
\colhead{} & 
\colhead{}
}
\startdata
B0040-2655 & 00 40 55.17 & -26 55 31.4 & 00 43 22.76 & -26 39 06.1 & 0.999 & 17.40 & SGP & \tablenotemark{a} \\
B0049-3024 & 00 49 43.78 & -30 24 12.4 & 00 52 08.94 & -30 07 54.8 & 0.471 & 17.75 & SGP & \tablenotemark{b} \\
B0052+0148 & 00 52 55.70 & +01 48 13.7 & 00 55 29.99 & +02 04 28.1 & 0.595 & 18.66 & F826 & \nodata \\
B0059-2853 & 00 59 10.62 & -28 53 37.6 & 01 01 34.83 & -28 37 29.9 & 0.620 & 18.44 & SGP & \tablenotemark{c} \\
B0303-0132B & 03 03 16.80 & -01 32 16.4 & 03 05 49.35 & -01 20 41.9 & 0.604 & 17.93 & F832 & \nodata \\
B0110-0012 & 01 10 21.20 & -00 12 40.7 & 01 12 54.93 & +00 03 13.0 & 0.235 & 18.19 & F826 & \nodata \\
B1027-0149 & 10 27 13.99 & -01 49 29.1 & 10 29 46.86 & -02 04 52.6 & 0.755 & 18.33 & F854 & \nodata \\
B1217+0945 & 12 17 55.19 & +09 45 06.8 & 12 20 28.01 & +09 28 28.0 & 1.078 & 18.54 & VSW & \nodata \\
B1237-0054 & 12 37 25.85 & -00 54 16.6 & 12 39 59.82 & -01 10 44.6 & 0.820 & 18.11 & F861 & \tablenotemark{d} \\
B1247+0036 & 12 47 56.84 & +00 36 19.4 & 12 50 30.48 & +00 20 00.1 & 1.155 & 18.23 & F861 & \tablenotemark{e} \\
B2132-4228 & 21 32 15.07 & -42 28 28.0 & 21 35 25.11 & -42 15 02.8 & 0.569 & 17.18 & F287 & \nodata \\
B2132-4516 & 21 32 13.39 & -45 16 50.3 & 21 35 27.17 & -45 03 25.0 & 0.507 & 17.69 & F287 & \tablenotemark{f} \\

\enddata

\tablenotetext{a}{Obvious candidate. Observed in 1991 November but not included in Paper VI due to
                    book keeping error. See also Cristiani et al. (1995)}
\tablenotetext{b}{Obvious candidate. Not observed due to erroneous information, via Private Communication,
                    that objects was a star. Redshift from Cristiani et al. (1995)}
\tablenotetext{c}{Candidate. Redshift from Warren et al. (1991)}
\tablenotetext{d}{Candidate. See also Boyle et al. (1990)}
\tablenotetext{e}{Obvious candidate. Not in Paper VI sample due to book keeping error (in Goldschmidt 1993)}
\tablenotetext{f}{Obvious candidate. Redshift from Hawkins and Veron (1995)}

\end{deluxetable}

\begin{deluxetable}{lcccccrcccl}
%\rotate
\tabletypesize{\scriptsize}
\tablewidth{0pt}
\tablenum{2}
\tablecaption{LBQS/FIRST Quasars}

\tablehead{
\colhead{LBQS Name} &
\colhead{R.A.} & 
\colhead{Dec.} & 
\colhead{$z$} &
\colhead{$B_J$} &
\colhead{FIRST Source} & 
\colhead{$S_i$} &
\colhead{$\sigma_i$} &
\colhead{Separation} & 
\colhead{$LogR^*$} & 
\colhead{Notes}
\\
\colhead{} & 
\colhead{(J2000.0)} & 
\colhead{(J2000.0)} & 
\colhead{} & 
\colhead{} & 
\colhead{} &
\colhead{(mJy)} & 
\colhead{(mJy)} & 
\colhead{(arcsec)} & 
\colhead{} &
\colhead{}
}
\startdata

B2359-0021 & 00 02 22.48 & -00 04 43.4 & 0.810 & 18.56 & J000222.4-000443    & 3.89 & 0.151 & 0.1  & 1.23 & \\
B0002-0149 & 00 05 07.06 & -01 32 45.2 & 1.710 & 18.72 & J000507.0-013245   & 63.87 & 0.145 & 0.1  & 2.42 & \\
B0004+0036 & 00 07 10.00 & +00 53 28.9 & 0.317 & 17.79 & J000710.0+005328    & 1.44 & 0.115 & 0.4  & 0.56 & \\
B0009-0148 & 00 12 14.83 & -01 31 28.7 & 1.072 & 17.77 & J001214.8-013128    & 1.08 & 0.149 & 0.4  & 0.33 & \tablenotemark{ns} \\
B0012-0016 & 00 14 44.03 & -00 00 18.6 & 1.559 & 18.21 & J001444.0-000017    & 1.51 & 0.138 & 0.9  & 0.60 & \\
B0012-0024 & 00 15 07.00 & -00 08 00.9 & 1.701 & 18.65 & J001507.0-000800   & 13.53 & 0.151 & 0.3  & 1.72 & \\
B0019+0022A & 00 21 41.04 & +00 38 41.5 & 0.314 & 18.64 & J002141.0+003841    & 1.38 & 0.171 & 0.2 & 0.88 & \\
B0020-0202 & 00 22 44.30 & -01 45 51.0 & 0.691 & 18.37 & J002244.2-014551    & 9.48 & 0.152 & 0.0  & 1.55 & \\
B0021-0100 & 00 24 11.66 & -00 43 47.8 & 0.771 & 18.18 & J002411.6-004349    & 0.99 & 0.143 & 1.4  & 0.49 & \\
B0024+0020 & 00 27 17.38 & +00 37 23.2 & 1.228 & 17.97 & J002717.3+003723    & 3.73 & 0.152 & 0.7  & 0.93 & \\
B0029-0152 & 00 31 36.48 & -01 36 21.6 & 2.383 & 18.65 & J003136.4-013622   & 14.06 & 0.138 & 0.4  & 1.69 & \\
B0048+0025 & 00 51 30.48 & +00 41 50.0 & 1.188 & 18.16 & J005130.4+004149   & 13.78 & 0.152 & 0.2  & 1.58 & \\
B0049+0019A & 00 52 05.55 & +00 35 38.2 & 0.399 & 16.46 & J005205.5+003538   & 87.95 & 0.140 & 0.2 & 1.80 & \\
B0052-0015 & 00 54 41.22 & +00 01 10.5 & 0.648 & 17.72 & J005441.1+000110    & 2.69 & 0.145 & 0.6  & 0.75 & \tablenotemark{ns} \\
B0056-0009 & 00 59 05.52 & +00 06 51.4 & 0.717 & 17.73 & J005905.5+000651 & 2415.95 & 0.409 & 0.3  & 3.70 & \\
B0059-0206 & 01 02 05.60 & -01 50 38.5 & 1.321 & 18.00 & J010205.6-015038    & 1.81 & 0.150 & 0.4  & 0.62 & \\
B0107-0235 & 01 10 13.17 & -02 19 52.5 & 0.958 & 18.11 & J011013.2-021950  & 147.45 & 0.138 & 2.0  & 2.61 & \\
B0249+0044 & 02 51 56.30 & +00 57 06.6 & 0.470 & 18.46 & J025156.3+005706   &  8.14 & 0.150 & 0.3  & 1.55 & \tablenotemark{ns} \\
B0251-0001 & 02 53 40.93 & +00 11 10.0 & 1.682 & 18.36 & J025340.9+001110    & 7.64 & 0.149 & 0.5  & 1.36 & \\
B0256-0206 & 02 58 38.06 & -01 54 10.7 & 0.406 & 18.46 & J025838.0-015411    & 1.14 & 0.131 & 0.7  & 0.71 & \\
B0256-0000 & 02 59 05.66 & +00 11 22.3 & 3.364 & 18.22 & J025905.6+001122    & 2.59 & 0.141 & 0.7  & 0.72 & \\
B0256-0031 & 02 59 28.54 & -00 19 59.7 & 1.995 & 17.59 & J025928.5-001959  & 224.97 & 0.275 & 0.4  & 2.49 & \\
 & & & &                                               & J025927.9-002004    & 9.63 & 0.275 & 9.8  &      & \\
 & & & &                                               & J025929.0-001956   & 12.69 & 0.276 & 8.4  &      & \\
B0257+0025 & 02 59 37.48 & +00 37 36.5 & 0.532 & 16.81 & J025937.4+003735    & 0.78 & 0.152 & 0.7  & -0.13 & \\
B1012+0213 & 10 15 15.67 & +01 58 52.8 & 1.378 & 17.63 & J101515.5+015851  & 213.47 & 0.161 & 2.6  & 2.54 & \tablenotemark{N} \\
 & & & &                                               & J101516.0+015852  & 516.70 & 0.161 & 5.1  &      & \\
B1013+0124 & 10 15 57.08 & +01 09 13.6 & 0.779 & 16.62 & J101557.0+010913  & 132.66 & 0.147 & 0.4  & 1.99 & \\
B1015+0147 & 10 17 42.40 & +01 32 17.2 & 1.455 & 18.30 & J101742.3+013217    & 6.33 & 0.154 & 0.4  & 1.27 & \\
B1016-0248 & 10 19 00.86 & -03 03 50.3 & 0.717 & 18.46 & J101900.8-030350    & 3.29 & 0.156 & 0.2  & 1.13 & \\
B1017-0009 & 10 19 56.77 & -00 24 10.0 & 1.127 & 17.50 & J101956.7-002409    & 8.05 & 0.156 & 0.4  & 1.09 & \\
B1025+0145 & 10 28 15.96 & +01 30 06.5 & 1.055 & 18.39 & J102815.8+013006   & 69.79 & 0.136 & 2.0  & 2.39 & \\
B1026-0045B & 10 28 37.02 & -01 00 27.4 & 1.530 & 18.42 & J102837.0-010027  & 156.41 & 0.144 & 0.2  & 2.70 & \\
B1130+0032 & 11 33 03.00 & +00 15 48.8 & 1.173 & 18.62 & J113303.0+001548  & 224.21 & 0.149 & 0.7  & 2.97 & \\
B1137-0048 & 11 40 04.35 & -01 05 27.5 & 0.347 & 18.00 & J114004.3-010527    & 9.73 & 0.142 & 0.2  & 1.47 & \\
B1137+0110 & 11 40 16.70 & +00 53 51.4 & 1.138 & 18.20 & J114016.7+005351  & 160.30 & 0.388 & 0.5  & 2.66 & \\
B1138+0003 & 11 41 17.64 & -00 12 50.4 & 0.500 & 17.90 & J114117.6-001250   & 16.73 & 0.153 & 0.3  & 1.64 & \\
B1148-0007 & 11 50 43.88 & -00 23 54.5 & 1.976 & 17.25 & J115043.8-002354 & 2816.39 & 0.635 & 0.6  & 3.46 & \\
 & & & &                                               & J115044.1-002354   & 58.78 & 0.635 & 4.1  &      & \\
B1148-0033 & 11 50 52.29 & -00 50 16.5 & 0.800 & 17.69 & J115052.2-005016    & 0.99 & 0.164 & 0.7  & 0.29 & \\
B1229-0207 & 12 32 00.03 & -02 24 04.8 & 1.045 & 17.65 & J123159.9-022405 & 1381.17 & 0.172 & 1.2  & 3.39 & \\
B1230-0015 & 12 33 04.02 & -00 31 34.2 & 0.470 & 17.00 & J123304.0-003134   & 75.67 & 0.141 & 0.5  & 1.94 & \\
B1234-0212 & 12 36 39.87 & -02 28 35.9 & 0.305 & 18.01 & J123639.8-022836    & 1.16 & 0.152 & 0.7  & 0.55 & \\
B1236-0043 & 12 38 56.12 & -00 59 30.9 & 1.843 & 18.46 & J123856.1-005930   & 14.04 & 0.145 & 0.2  & 1.65 & \\
B1240+0224 & 12 42 47.17 & +02 08 17.4 & 0.790 & 17.92 & J124247.1+020816   & 15.56 & 0.154 & 1.3  & 1.58 & \\
B1243-0026 & 12 46 13.12 & -00 42 33.0 & 0.644 & 17.00 & J124613.1-004232    & 2.15 & 0.140 & 0.4  & 0.37 & \\
B1308-0111 & 13 10 55.58 & -01 27 25.3 & 1.004 & 18.71 & J131055.5-012724    & 4.56 & 0.142 & 0.5  & 1.34 & \\
B1308+0051 & 13 11 06.46 & +00 35 10.1 &   ?   & 18.65 & J131106.4+003510   & 22.75 & 0.151 & 0.1  & 2.04 & \tablenotemark{a} \\
B1308-0214 & 13 11 14.92 & -02 30 46.0 & 2.885 & 18.68 & J131114.9-023045   & 55.40 & 0.155 & 0.4  & 2.26 & \\
B1308+0109 & 13 11 21.15 & +00 53 19.6 & 1.075 & 18.07 & J131121.1+005319  & 144.57 & 0.156 & 0.5  & 2.57 & \\
B1310+0216 & 13 13 30.12 & +02 01 05.5 &   ?   & 18.60 & J131330.1+020105   & 52.89 & 0.139 & 0.3  & 2.39 & \tablenotemark{a} \\
B1313-0142 & 13 15 38.71 & -01 58 46.0 & 1.498 & 18.79 & J131538.5-015846  & 113.11 & 0.157 & 3.2  & 2.71 & \tablenotemark{N} \\ 
 & & & &                                               & J131539.0-015845  & 155.44 & 0.157 & 4.5  &      & \\
B1313+0107 & 13 16 30.46 & +00 51 25.5 & 2.393 & 18.13 & J131630.4+005125    & 2.20 & 0.135 & 0.1  & 0.67 & \\
B1314-0216 & 13 17 26.12 & -02 31 50.6 & 1.090 & 18.52 & J131726.1-023150   & 22.04 & 0.149 & 0.4  & 1.93 & \\
B1317-0033 & 13 19 38.76 & -00 49 39.6 & 0.892 & 18.21 & J131938.7-004939 & 1536.50 & 0.422 & 0.6  & 3.68 & \\
B1317-0142 & 13 19 50.35 & -01 58 03.7 & 0.225 & 17.26 & J131950.3-015803    & 1.38 & 0.138 & 0.3  & 0.34 & \\
B1319+0039 & 13 21 39.57 & +00 23 57.9 & 1.619 & 17.89 & J132139.7+002356  & 179.01 & 0.148 & 3.0  & 2.54 & \tablenotemark{N} \\
B1323-0248 & 13 26 15.16 & -03 03 57.9 & 2.120 & 17.36 & J132615.1-030357    & 5.48 & 0.155 & 0.6  & 0.78 & \\
B1324+0039 & 13 27 12.84 & +00 24 14.4 & 1.061 & 18.22 & J132712.8+002414   & 84.27 & 0.136 & 0.6  & 2.40 & \\
B1325+0027 & 13 27 50.43 & +00 11 57.0 & 2.534 & 18.72 & J132750.4+001156   & 24.99 & 0.147 & 0.7  & 1.96 & \\
B1331-0108 & 13 34 28.05 & -01 23 48.8 & 1.881 & 17.87 & J133428.0-012349    & 2.99 & 0.144 & 0.3  & 0.74 & \\
B1331-0123 & 13 34 33.23 & -01 38 25.2 & 0.290 & 18.32 & J133433.2-013825    & 1.80 & 0.145 & 0.6  & 0.87 & \tablenotemark{ns} \\
B1332-0045 & 13 35 26.00 & -01 00 28.3 & 0.672 & 17.41 & J133526.0-010026    & 1.00 & 0.152 & 1.7  & 0.20 & \\
B1335+0222 & 13 37 39.65 & +02 06 57.1 & 1.354 & 18.36 & J133739.6+020657  & 194.94 & 0.151 & 0.3  & 2.79 & \\
B1338-0038 & 13 41 13.93 & -00 53 15.1 & 0.236 & 17.91 & J134113.9-005314    & 5.35 & 0.153 & 0.3  & 1.19 & \\
B1429-0053 & 14 32 29.27 & -01 06 16.2 & 2.078 & 17.68 & J143229.3-010614    & 0.93 & 0.161 & 1.6  & 0.14 & \\
B1430-0046 & 14 32 44.41 & -00 59 15.3 & 1.023 & 17.75 & J143244.4-005914   & 16.53 & 0.145 & 1.0  & 1.51 & \\
B1438+0210 & 14 40 59.48 & +01 57 44.2 & 0.797 & 18.37 & J144059.4+015743  & 214.21 & 0.142 & 0.3  & 2.90 & \\
B1443-0100 & 14 45 59.51 & -01 13 17.3 & 1.794 & 18.30 & J144559.5-011317    & 2.79 & 0.154 & 1.3  & 0.89 & \\
B1446-0035 & 14 49 30.51 & -00 47 46.0 & 0.254 & 18.07 & J144930.5-004745    & 2.36 & 0.145 & 0.5  & 0.90 & \tablenotemark{ns} \\
B2231-0048 & 22 33 59.96 & -00 33 15.3 & 1.209 & 17.57 & J223400.0-003315    & 1.03 & 0.145 & 1.2  & 0.21 & \\
B2235+0054 & 22 37 34.17 & +01 10 34.6 & 0.529 & 18.55 & J223734.1+011035    & 1.14 & 0.136 & 1.1  & 0.73 & \\
B2245-0055 & 22 47 39.31 & -00 39 52.6 & 0.801 & 17.43 & J224739.3-003953    & 1.58 & 0.155 & 1.1  & 0.39 & \\
B2351-0036 & 23 54 09.16 & -00 19 48.1 & 0.460 & 18.47 & J235409.1-001947  & 346.10 & 0.140 & 0.5  & 3.19 & \\

\enddata

\tablenotetext{N}{Optical--radio separation $> 2.1\,$arcsec. Object not included in LBQS--FIRST statistical sample.}
\tablenotetext{ns}{Classified as a non--stellar source on the UKST direct plate}
\tablenotetext{a}{BL Lac candidate without confirmed redshift. Redshift z=0.75 adopted in calculation of $LogR^*$}

\end{deluxetable}

\begin{deluxetable}{ccccrrcclcl}
%\rotate
\tabletypesize{\scriptsize}
\tablewidth{0pt}
\tablenum{3}
\tablecaption{LBQS/FIRST Stellar Matches}

\tablehead{
\colhead{R.A.} & 
\colhead{Dec.} & 
\colhead{$B_J$} &
\colhead{FIRST Source} & 
\colhead{$S_i$} &
\colhead{$\sigma_i$} &
\colhead{Separation} & 
\colhead{LBQS Sp.} &
\colhead{Class Sp.} &
\colhead{$z$} &
\colhead{Source}
\\
\colhead{(J2000.0)} & 
\colhead{(J2000.0)} & 
\colhead{} & 
\colhead{} &
\colhead{(mJy)} & 
\colhead{(mJy)} & 
\colhead{(arcsec)} & 
\colhead{} &
\colhead{} &
\colhead{} &
\colhead{}
}
\startdata
00 13 01.28 & -00 27 12.3 & 18.24 & J001301.3-002714 &   3.58 & 0.151 & 2.1  & \nodata & Star    & \nodata & Becker \\
00 13 56.17 & -01 42 29.8 & 16.35 & J001356.2-014229 &   6.08 & 0.150 & 1.9  & \nodata & Star    & \nodata & Becker \\
00 17 01.43 & -00 13 09.7 & 17.75 & J001701.3-001308 &   2.04 & 0.189 & 1.7  & \nodata & Star    & \nodata & Becker \\
01 03 52.47 & +00 37 39.9 & 17.54 & J010352.4+003739 &   2.22 & 0.143 & 0.2  & \nodata & Quasar  & 0.704   & Becker \\
03 06 56.29 & +00 44 31.5 & 16.72 & J030656.2+004431 &   0.79 & 0.143 & 0.7  & \nodata & Star    & \nodata & Becker \\
10 22 43.73 & -01 13 02.1 & 16.54 & J102243.7-011302 &  37.96 & 0.183 & 0.2  & QSO?    & BL Lac   & \nodata & LBQS \\
10 32 42.44 & +02 02 28.9 & 18.67 & J103242.5+020229 &   1.73 & 0.148 & 1.4  & QSO?    & Quasar  & 0.537   & Keck \\
11 43 54.02 & +01 13 43.3 & 17.54 & J114354.0+011343 &   0.80 & 0.162 & 0.5  & QSO?    & Quasar  & 1.282   & Keck \\
11 46 31.44 & -02 20 42.4 & 15.99 & J114631.3-022042 &   1.38 & 0.141 & 1.1  & \nodata & Star    & \nodata & Keck \\
12 40 35.84 & -00 29 19.6 & 17.74 & J124035.7-002920 &   1.29 & 0.150 & 1.9  & \nodata & AGN     & 0.081   & Keck \\
12 50 06.82 & +01 58 03.8 & 18.45 & J125006.8+015803 & 155.93 & 0.153 & 0.7  & QSO?    & Quasar  & 0.941   & Keck \\
13 13 27.48 & -02 32 32.8 & 18.03 & J131327.4-023232 &   1.67 & 0.148 & 0.2  & QSO??   & AGN     & 0.174   & Keck \\
13 16 37.25 & -00 36 35.8 & 18.37 & J131637.2-003635 &   1.94 & 0.135 & 0.3  & QSO??   & Quasar  & 0.895   & Keck \\
13 22 11.94 & +01 30 34.5 & 17.58 & J132211.9+013034 &   1.01 & 0.135 & 0.3  & QSO??   & AGN     & 0.190   & LBQS \\
13 33 56.01 & +00 12 29.1 & 17.64 & J133356.0+001229 &   2.07 & 0.141 & 0.9  & QSO?    & Quasar  & 0.918   & Keck \\
13 36 16.84 & -00 15 14.2 & 16.61 & J133616.8-001515 &  65.68 & 0.152 & 1.6  & \nodata & Star    & \nodata & Keck \\
14 38 06.79 & +00 32 20.1 & 18.33 & J143806.7+003220 &   2.30 & 0.152 & 0.7  & \nodata & AGN & 0.209   & Keck \\
14 50 13.18 & +00 35 35.3 & 18.01 & J145013.1+003535 &   8.16 & 0.151 & 0.9  & \nodata & Star    & \nodata & Keck \\
22 37 23.97 & -00 09 53.3 & 16.79 & J223724.0-000953 &   9.54 & 0.135 & 1.8  & \nodata & Star    & \nodata & Becker \\
22 43 53.32 & +00 03 18.8 & 17.99 & J224353.1+000318 &   4.34 & 0.149 & 1.8  & \nodata & Star    & \nodata & MMT \\
22 51 21.11 & +01 24 13.1 & 17.95 & J225121.1+012414 &   1.14 & 0.149 & 1.5  & \nodata & Star    & \nodata & Becker \\

\enddata

\end{deluxetable}

\begin{deluxetable}{ccccrcccl}
\tabletypesize{\scriptsize}
\tablewidth{0pt}
\tablenum{4}
\tablecaption{LBQS/FIRST Non-Stellar Matches}

\tablehead{
\colhead{R.A.} & 
\colhead{Dec.} & 
\colhead{$B_J$} &
\colhead{FIRST Source} & 
\colhead{$S_i$} &
\colhead{$\sigma_i$} &
\colhead{Separation} & 
\colhead{$z$} &
\colhead{Notes}
\\
\colhead{(J2000.0)} & 
\colhead{(J2000.0)} & 
\colhead{} & 
\colhead{} &
\colhead{(mJy)} & 
\colhead{(mJy)} & 
\colhead{(arcsec)} & 
\colhead{} &
\colhead{}
}
\startdata

 00 02 58.62 & +00 08 30.8 & (16.2) & J000258.5+000830 & 0.84 & 0.146 & 0.7 & 0.090 & MMT \\ 
 00 03 32.27 & +00 52 58.2 & (16.0) & J000332.2+005257 & 1.42 & 0.130 & 0.7 & & \\
 00 04 58.56 & +00 51 34.5 & (18.1) & J000458.6+005135 & 0.93 & 0.127 & 1.3 & & \\
 00 05 21.22 & -01 32 30.4 & (16.7) & J000521.2-013230 & 1.76 & 0.152 & 0.5 & 0.110 & MMT \\
 00 08 13.22 & -00 57 53.2 & (17.2) & J000813.1-005751 & 2.74 & 0.150 & 1.6 & 0.139 & MMT \\
 00 11 27.08 & +00 09 32.4 & (16.2) & J001127.0+000932 & 0.77 & 0.153 & 0.1 & 0.108 & MMT \\
 00 11 55.47 & +01 14 28.8 & (16.9) & J001155.4+011428 & 4.40 & 0.134 & 0.6 & & \\
 00 12 26.85 & -00 48 19.3 & (16.2) & J001226.8-004819 & 5.31 & 0.139 & 0.3 & 0.073 & \tablenotemark{a} \\
 00 12 47.57 & +00 47 15.9 & (16.6) & J001247.6+004715 & 5.46 & 0.143 & 1.7 & & \\
 00 15 08.44 & +00 13 16.7 & (17.5) & J001508.4+001315 & 0.72 & 0.154 & 0.9 & & \\
 00 16 13.09 & +00 20 06.4 & (16.1) & J001613.0+002006 & 1.43 & 0.142 & 0.8 & & \\
 00 16 30.41 & +01 22 05.1 & (17.6) & J001630.4+012205 & 2.37 & 0.125 & 0.0 & & \\
 00 20 16.13 & +00 04 46.1 & (17.5) & J002016.1+000445 & 55.02 & 0.203 & 0.3 & & \\
 00 20 43.95 & -00 26 23.7 & (16.5) & J002043.9-002624 & 1.11 & 0.135 & 1.3 & & \\
 00 21 06.47 & -01 39 20.3 & (16.9) & J002106.4-013920 & 1.83 & 0.138 & 0.9 & 0.084 & MMT \\
 00 22 56.88 & +01 15 55.6 & (17.1) & J002256.8+011557 & 2.20 & 0.127 & 1.4 & & \\
 00 23 45.09 & -01 00 29.4 & (16.6) & J002345.0-010029 & 3.01 & 0.145 & 0.6 & 0.067 & \tablenotemark{a} \\
 00 25 02.85 & -02 14 46.3 & (17.7) & J002502.8-021445 & 13.08 & 0.148 & 0.4 & & \\
 00 25 26.33 & -00 55 12.5 & (17.4) & J002526.3-005512 & 4.48 & 0.148 & 0.5 & 0.139 & MMT \\
 00 26 48.40 & +01 17 37.7 & (17.3) & J002648.3+011737 & 2.10 & 0.133 & 0.3 & & \\
 00 28 33.41 & +00 55 10.8 & (16.6) & J002833.4+005511 & 237.22 & 0.115 & 0.3 & 0.105 & \tablenotemark{b} \\
 00 29 29.54 & -00 09 13.1 & (17.7) & J002929.5-000911 & 1.06 & 0.164 & 1.5 & 0.168 & MMT \\
 00 32 30.92 & -00 24 40.1 & (16.1) & J003230.9-002440 & 27.74 & 0.141 & 0.1 & 0.081 & MMT \\
 00 50 59.37 & -00 14 54.0 & (16.4) & J005059.4-001453 & 2.95 & 0.147 & 0.7 & 0.135 & MMT \\
 00 52 52.38 & -00 04 27.7 & (16.1) & J005252.3-000426 & 1.64 & 0.146 & 1.1 & 0.120 & \tablenotemark{c} \\
 00 53 25.47 & -02 12 29.8 & (17.5) & J005325.4-021231 & 3.00 & 0.145 & 1.6 & 0.117 & MMT \\
 00 55 28.82 & +00 34 52.5 & (16.1) & J005528.8+003452 & 1.44 & 0.134 & 0.5 & & \\
 00 55 38.31 & -02 11 27.1 & (16.7) & J005538.3-021127 & 3.12 & 0.157 & 0.4 & & \\
 00 56 47.87 & +00 04 26.0 & (17.5) & J005647.8+000427 & 1.43 & 0.137 & 2.0 & 0.116 & MMT \\
 00 59 16.93 & -01 50 17.7 & (17.5) & J005916.9-015017 & 13.38 & 0.163 & 0.1 & & \\
 01 00 32.22 & -02 00 46.3 & (18.0) & J010032.2-020046 & 5.95 & 0.152 & 0.1 & 0.227 & MMT \\
 01 04 00.87 & -00 19 16.9 & (16.1) & J010400.8-001916 & 1.51 & 0.148 & 0.5 & 0.051 & MMT \\
 01 05 25.13 & +00 57 57.6 & (16.4) & J010525.0+005757 & 0.84 & 0.146 & 0.9 & & \\
 01 06 22.65 & +00 58 51.9 & (16.2) & J010622.6+005852 & 1.92 & 0.143 & 0.4 & & \\
 01 06 24.86 & -01 41 47.2 & (17.1) & J010624.9-014147 & 0.53 & 0.139 & 0.5 & & \\
 01 06 40.53 & -00 02 35.4 & (17.0) & J010640.4-000235 & 0.80 & 0.148 & 0.9 & & \\
 01 09 39.02 & +00 59 50.6 & (16.7) & J010939.0+005949 & 1.19 & 0.166 & 0.8 & & \\
 01 10 36.24 & -00 28 06.2 & (16.4) & J011036.2-002805 & 2.49 & 0.149 & 0.6 & 0.110 & MMT \\
 02 55 57.12 & -01 16 39.9 & (17.2) & J025557.2-011639 & 1.51 & 0.153 & 1.7 & 0.080 & MMT \\
 03 00 15.98 & +01 18 50.6 & (17.6) & J030015.9+011850 & 1.46 & 0.141 & 0.5 & 0.155 & MMT \\
 03 03 26.86 & -02 02 10.3 & (16.3) & J030326.8-020209 & 3.62 & 0.146 & 0.8 & 0.119 & MMT \\
 03 03 35.01 & +00 44 40.0 & (17.7) & J030335.0+004439 & 1.06 & 0.149 & 1.1 & & \\
 03 08 32.68 & -02 15 28.8 & (16.5) & J030832.6-021529 & 2.31 & 0.151 & 0.7 & 0.121 & MMT \\
 03 09 15.57 & +00 25 08.3 & (16.2) & J030915.5+002508 & 2.82 & 0.132 & 0.8 & & \\
 03 09 48.07 & -02 16 02.5 & (17.3) & J030948.0-021603 & 4.61 & 0.153 & 1.0 & 0.107 & MMT \\
 10 10 44.51 & +00 43 31.7 & (16.5) & J101044.5+004331 & 0.77 & 0.139 & 0.7 & 0.180 & LBQS \\
 10 11 41.08 & -01 50 44.2 & (17.7) & J101141.1-015043 & 0.74 & 0.152 & 1.4 & & \\
 10 12 02.08 & +00 52 52.4 & (18.0) & J101202.1+005251 & 1.08 & 0.136 & 1.4 & 0.123 & 2dFGRS \\
 10 15 32.75 & -00 45 28.4 & (16.1) & J101532.7-004528 & 1.57 & 0.152 & 0.2 & 0.063 & 2dFGRS \\
 10 15 36.25 & +00 54 59.3 & (17.1) & J101536.2+005458 & 2.50 & 0.151 & 0.7 & 0.121 & 2dFGRS \\
 10 16 15.92 & -03 02 46.2 & (16.7) & J101615.9-030246 & 0.80 & 0.156 & 0.4 & 0.158 & 2dFGRS \\
 10 17 17.94 & +01 06 56.6 & (17.9) & J101717.9+010656 & 0.82 & 0.154 & 0.4 & 0.078 & 2dFGRS \\
 10 17 33.23 & -00 01 45.5 & (16.3) & J101733.2-000145 & 2.19 & 0.139 & 0.4 & & \\
 10 20 18.32 & +01 27 11.3 & (17.2) & J102018.3+012712 & 0.93 & 0.153 & 1.0 & 0.115 & 2dFGRS \\
 10 21 00.09 & +00 07 50.9 & (17.2) & J102100.0+000751 & 5.15 & 0.142 & 0.4 & 0.112 & 2dFGRS \\
 10 21 08.22 & -00 51 19.1 & (16.4) & J102108.1-005119 & 1.13 & 0.142 & 1.1 & & \\
 10 22 17.84 & -01 55 59.3 & (17.7) & J102217.8-015559 & 2.92 & 0.144 & 0.5 & & \\
 10 23 09.04 & -01 55 37.8 & (16.5) & J102309.0-015538 & 8.83 & 0.155 & 0.8 & & \\
 10 24 01.47 & +00 00 38.8 & (16.5) & J102401.5+000039 & 6.01 & 0.132 & 0.7 & 0.127 & 2dFGRS \\
 10 24 27.48 & -02 53 20.9 & (16.8) & J102427.5-025321 & 2.97 & 0.151 & 0.6 & 0.164 & \tablenotemark{d} \\
 10 26 07.01 & +01 31 45.4 & (18.2) & J102607.0+013145 & 26.17 & 0.144 & 0.8 & & \\
 10 26 35.17 & -00 01 58.0 & (17.5) & J102635.2-000157 & 1.41 & 0.141 & 1.0 & 0.110 & 2dFGRS \\
 10 27 23.27 & -02 32 28.8 & (18.3) & J102723.2-023227 & 16.30 & 0.143 & 0.9 & & \\
 10 27 35.27 & -02 16 12.7 & (18.2) & J102735.3-021612 & 4.06 & 0.152 & 0.5 & 0.218 & 2dFGRS \\
 10 28 18.60 & -01 22 07.8 & (16.0) & J102818.6-012208 & 0.71 & 0.176 & 1.3 & 0.127 & 2dFGRS \\
 10 31 56.77 & +01 30 45.3 & (17.2) & J103156.8+013046 & 7.98 & 0.138 & 1.1 & & \\
 11 30 34.67 & +00 52 51.3 & (16.8) & J113034.6+005251 & 7.83 & 0.148 & 0.6 & & \\
 11 31 22.66 & +01 08 05.9 & (17.2) & J113122.6+010805 & 15.81 & 0.139 & 0.5 & & \\
 11 32 37.03 & +01 49 14.8 & (17.1) & J113236.9+014914 & 0.85 & 0.143 & 0.4 & & \\
 11 32 38.83 & -01 31 48.7 & (17.0) & J113238.8-013147 & 1.86 & 0.156 & 1.0 & 0.100 & 2dFGRS \\
 11 32 52.17 & +01 18 28.2 & (17.6) & J113252.1+011830 & 1.74 & 0.138 & 1.8 & & \\
 11 32 58.74 & +00 55 28.3 & (17.1) & J113258.7+005528 & 2.97 & 0.141 & 0.3 & & \\
 11 37 45.33 & -02 51 53.9 & (16.1) & J113745.2-025152 & 2.60 & 0.141 & 2.0 & & \\
 11 39 41.62 & -01 04 59.7 & (16.8) & J113941.5-010500 & 68.85 & 0.155 & 1.0 & & \\
 11 42 36.96 & +01 55 53.4 & (16.7) & J114236.9+015552 & 2.88 & 0.185 & 0.7 & 0.120 & LBQS \\
 11 43 06.72 & +01 01 15.8 & (17.3) & J114306.7+010115 & 3.95 & 0.171 & 0.3 & & \\
 11 43 40.09 & -01 26 27.6 & (17.7) & J114340.1-012628 & 1.25 & 0.140 & 1.0 & 0.124 & 2dFGRS \\
 11 43 43.01 & -02 14 06.7 & (16.5) & J114343.0-021406 & 2.10 & 0.149 & 0.7 & 0.121 & 2dFGRS \\
 11 44 04.80 & -02 02 26.9 & (16.0) & J114404.8-020226 & 1.71 & 0.150 & 0.7 & 0.068 & 2dFGRS \\
 11 44 48.40 & -01 09 30.7 & (16.5) & J114448.4-010930 & 17.98 & 0.148 & 0.6 & & \\
 11 45 13.87 & -02 59 39.1 & (16.6) & J114513.8-025939 & 108.93 & 0.144 & 0.2 & 0.107 & 2dFGRS \\
 12 33 37.65 & -00 22 28.3 & (16.8) & J123337.6-002228 & 0.91 & 0.151 & 0.3 & & \\
 12 35 27.12 & -02 29 02.8 & (16.7) & J123527.1-022903 & 0.64 & 0.150 & 0.4 & 0.081 & 2dFGRS \\
 12 37 12.02 & +02 03 09.3 & (17.4) & J123712.0+020309 & 2.02 & 0.145 & 0.1 & & \\
 12 38 29.90 & -02 28 18.9 & (16.5) & J123829.9-022819 & 0.78 & 0.149 & 0.5 & 0.084 & 2dFGRS \\
 12 38 49.06 & -01 35 55.0 & (17.9) & J123849.1-013553 & 6.87 & 0.145 & 1.6 & & \\
 12 39 51.26 & +01 01 12.7 & (17.3) & J123951.2+010112 & 3.48 & 0.158 & 0.6 & & \\
 12 46 08.66 & -02 25 51.3 & (16.4) & J124608.7-022551 & 1.29 & 0.141 & 0.5 & & \\
 12 46 17.50 & -00 16 57.8 & (16.1) & J124617.4-001657 & 1.06 & 0.139 & 0.5 & 0.127 & 2dFGRS \\
 13 10 42.81 & -01 55 01.3 & (16.4) & J131042.7-015500 & 3.90 & 0.140 & 0.6 & 0.082 & Keck  \\
 13 11 40.29 & -02 04 22.3 & (17.4) & J131140.2-020420 & 1.56 & 0.144 & 1.6 & 0.082 & Keck  \\
 13 16 44.95 & +00 23 13.5 & (16.9) & J131644.8+002312 & 0.86 & 0.147 & 1.2 & 0.160 & Keck  \\
 13 17 49.71 & -01 19 04.8 & (17.9) & J131749.6-011906 & 1.36 & 0.136 & 1.9 & 0.079 & Keck  \\
 13 19 29.86 & -01 08 43.1 & (17.1) & J131929.8-010843 & 6.04 & 0.570 & 0.1 & 0.140 & Keck  \\
 13 20 58.25 & -02 14 15.3 & (16.8) & J132058.2-021415 & 3.22 & 0.139 & 0.0 & 0.094 & Keck  \\
 13 21 00.50 & +01 16 09.7 & (17.5) & J132100.4+011609 & 11.08 & 0.138 & 0.0 & 0.148 & Keck  \\
 13 21 23.35 & +01 23 29.8 & (16.7) & J132123.3+012329 & 1.05 & 0.145 & 0.6 & 0.136 & Keck  \\
 13 21 46.04 & -00 01 51.4 & (16.7) & J132146.0-000151 & 0.86 & 0.153 & 0.5 & 0.155 & Keck  \\
 13 21 52.38 & -01 31 45.5 & (16.2) & J132152.4-013145 & 0.72 & 0.146 & 1.2 & 0.101 & Keck  \\
 13 22 32.04 & -01 42 50.2 & (16.9) & J132232.0-014250 & 1.64 & 0.150 & 0.6 & 0.084 & Keck  \\
 13 24 39.80 & +00 22 16.5 & (16.9) & J132439.8+002216 & 2.90 & 0.151 & 0.2 & 0.109 & Keck  \\
 13 25 06.33 & +01 51 44.8 & (16.0) & J132506.2+015145 & 2.85 & 0.141 & 1.0 & 0.058 & 2dFGRS \\
 13 26 32.19 & +00 28 00.8 & (17.5) & J132632.2+002800 & 1.15 & 0.141 & 0.6 & 0.086 & Keck  \\
 13 26 56.96 & +01 11 54.8 & (16.6) & J132656.9+011155 & 2.43 & 0.140 & 0.8 & 0.067 & Keck  \\
 13 28 32.56 & -02 33 21.2 & (17.6) & J132832.5-023320 & 0.97 & 0.150 & 0.6 & 0.184 & Keck  \\
 13 28 34.15 & -01 29 17.5 & (16.6) & J132834.1-012917 & 15.59 & 0.143 & 0.5 & & \\
 13 28 34.40 & -03 07 44.4 & (17.6) & J132834.3-030745 & 8.81 & 0.147 & 0.8 & 0.150 & LBQS \\
 13 28 38.94 & +00 32 53.8 & (16.9) & J132838.9+003253 & 5.06 & 0.140 & 0.3 & 0.105 & 2dFGRS \\
 13 29 23.48 & -03 15 02.5 & (16.2) & J132923.4-031502 & 0.97 & 0.158 & 0.2 & 0.076 & LBQS \\
 13 30 25.41 & +01 29 41.1 & (16.1) & J133025.4+012941 & 1.23 & 0.154 & 0.8 & 0.083 & Keck  \\
 13 31 55.14 & -02 51 45.0 & (16.9) & J133155.1-025143 & 1.92 & 0.144 & 1.5 & 0.084 & \tablenotemark{d} \\
 13 34 50.44 & +01 02 18.7 & (18.2) & J133450.4+010220 & 1.70 & 0.140 & 1.5 & 0.246 & Keck  \\
 13 35 48.04 & -03 04 02.7 & (17.5) & J133548.0-030402 & 1.95 & 0.140 & 0.4 & 0.113 & Keck  \\
 13 38 22.59 & +02 16 47.6 & (16.1) & J133822.5+021649 & 3.16 & 0.151 & 1.7 & 0.111 & Keck  \\
 13 38 30.72 & -00 42 05.7 & (16.1) & J133830.7-004205 & 1.97 & 0.154 & 0.3 & 0.072 & Keck  \\
 13 41 01.66 & -00 42 44.2 & (18.0) & J134101.6-004245 & 0.81 & 0.148 & 0.9 & 0.113 & Keck  \\
 13 42 12.21 & -00 17 37.7 & (16.1) & J134212.2-001737 & 1.50 & 0.146 & 0.1 & 0.087 & 2dFGRS \\
 13 43 05.89 & +00 48 24.7 & (16.8) & J134305.8+004824 & 2.02 & 0.151 & 0.2 & 0.145 & Keck  \\
 13 43 21.33 & -02 52 55.1 & (16.3) & J134321.3-025253 & 1.98 & 0.135 & 1.7 & 0.149 & \tablenotemark{d} \\
 13 44 28.33 & +00 01 46.6 & (16.7) & J134428.3+000146 & 8.01 & 0.146 & 0.3 & 0.135 & 2dFGRS \\
 13 45 29.89 & -01 40 02.2 & (16.1) & J134529.9-014003 & 3.98 & 0.145 & 1.1 & 0.088 & Keck  \\
 13 46 29.43 & +02 09 39.0 & (16.3) & J134629.4+020939 & 31.21 & 0.152 & 0.2 & 0.138 & Keck  \\
 13 49 28.08 & -00 45 07.5 & (16.3) & J134928.0-004508 & 0.75 & 0.142 & 0.9 & 0.143 & Keck  \\
 13 50 02.05 & +00 01 51.4 & (17.6) & J135002.0+000152 & 1.67 & 0.156 & 0.8 & 0.203 & 2dFGRS \\
 14 32 09.60 & -02 11 36.2 & (17.2) & J143209.6-021136 & 46.30 & 0.157 & 0.2 & 0.123 & 2dFGRS \\
 14 32 37.23 & +01 20 51.0 & (17.3) & J143237.2+012050 & 3.09 & 0.142 & 0.3 & 0.138 & Keck  \\
 14 33 28.36 & -00 02 01.5 &  16.6  & J143328.4-000202 & 2.85 & 0.139 & 1.1 & Star  & Keck  \\
 14 35 27.71 & +02 19 40.2 & (16.8) & J143527.7+021939 & 1.46 & 0.155 & 0.3 & 0.084 & Keck  \\
 14 35 46.97 & +02 01 55.1 & (17.1) & J143546.9+020154 & 0.79 & 0.146 & 0.4 & 0.120 & Keck  \\
 14 36 33.54 & -02 04 16.4 & (17.0) & J143633.5-020416 & 10.80 & 0.149 & 0.6 & 0.133 & Keck  \\
 14 37 43.38 & -00 27 45.8 & (17.1) & J143743.3-002746 & 2.30 & 0.154 & 1.0 & 0.141 & 2dFGRS \\
 14 38 37.80 & -00 22 50.0 &  18.2  & J143837.8-002251 & 1.27 & 0.143 & 1.9 & Star  & Keck  \\
 14 39 19.04 & +00 38 19.4 & (16.7) & J143919.0+003819 & 14.22 & 0.157 & 0.4 & 0.151 & Keck  \\
 14 40 14.62 & -02 46 09.8 & (16.8) & J144014.6-024609 & 3.35 & 0.138 & 0.3 & 0.130 & 2dFGRS \\
 14 47 07.50 & -02 08 44.8 & (16.3) & J144707.5-020845 & 2.79 & 0.156 & 0.5 & 0.071 & 2dFGRS \\
 14 47 24.02 & -00 24 11.4 & (16.9) & J144724.0-002411 & 3.36 & 0.163 & 0.8 & 0.175 & 2dFGRS \\
 14 47 26.28 & +00 50 16.1 & (17.0) & J144726.3+005016 & 2.35 & 0.147 & 0.7 & 0.172 & 2dFGRS \\
 14 49 28.61 & -01 16 16.8 & (17.0) & J144928.6-011615 & 40.16 & 0.151 & 1.0 & 0.202 & 2dFGRS \\
 14 50 08.30 & +01 51 59.3 & (16.0) & J145008.3+015200 & 3.86 & 0.148 & 0.9 & 0.113 & Keck  \\
 14 52 42.64 & +01 40 59.0 & (17.5) & J145242.7+014059 & 2.77 & 0.151 & 1.5 & 0.150 & Keck  \\
 14 53 05.74 & +01 55 01.2 & (17.6) & J145305.7+015501 & 2.58 & 0.150 & 0.9 & 0.098 & Keck  \\
 22 31 12.06 & -01 01 11.1 & (17.0) & J223112.0-010111 & 1.38 & 0.138 & 0.5 & & \\
 22 35 36.03 & -00 49 24.6 & (17.3) & J223536.0-004924 & 1.28 & 0.138 & 0.4 & 0.131 & MMT \\
 22 37 29.21 & -00 49 08.5 & (16.1) & J223729.2-004908 & 3.76 & 0.148 & 0.6 & 0.129 & MMT \\
 22 39 18.95 & -01 04 39.9 & (17.9) & J223918.9-010440 & 1.31 & 0.153 & 0.8 & 0.129 & MMT \\
 22 42 12.95 & -01 55 06.7 & (16.7) & J224212.8-015507 & 2.83 & 0.149 & 0.9 & 0.178 & MMT \\
 22 45 54.95 & +00 23 09.6 & (16.3) & J224554.9+002310 & 5.19 & 0.153 & 0.5 & & \\
 22 50 47.05 & -01 41 16.4 & (16.5) & J225047.0-014117 & 16.81 & 0.135 & 1.0 & 0.110 & MMT \\
 22 51 08.55 & -00 53 43.5 & (16.7) & J225108.6-005343 & 1.89 & 0.137 & 1.4 & 0.162 & MMT \\
 22 53 23.56 & +00 38 29.4 & (16.7) & J225323.6+003828 & 0.91 & 0.155 & 1.3 & & \\
 23 54 20.52 & +00 33 26.9 & (16.6) & J235420.5+003327 & 1.00 & 0.175 & 0.9 & & \\
 23 57 21.75 & +00 26 05.8 & (16.1) & J235721.7+002606 & 1.28 & 0.137 & 0.8 & & \\
 23 59 48.54 & -00 04 25.7 & (16.3) & J235948.5-000425 & 1.63 & 0.131 & 0.5 & 0.061 & MMT \\
\enddata

\tablenotetext{a}{Clements et al. (1996)}
\tablenotetext{b}{Wills and Wills (1974)}
\tablenotetext{c}{Suntzeff et al. (1996)}
\tablenotetext{d}{Shectman et al. (1996)}

\end{deluxetable}

\clearpage

\begin{deluxetable}{ccccccrccccl}
%\rotate
\tabletypesize{\scriptsize}
\tablewidth{0pt}
\tablenum{5}
\tablecaption{Additional LBQS/FIRST Quasars}

\tablehead{
\colhead{Name} &
\colhead{R.A.} & 
\colhead{Dec.} & 
\colhead{$z$} &
\colhead{$B_J$} &
\colhead{FIRST Source} & 
\colhead{$S_i$} &
\colhead{$\sigma_i$} &
\colhead{Separation} & 
\colhead{$LogR^*$} &
\colhead{LBQS Sp.} &
\colhead{Class Sp.}

\\
\colhead{} &
\colhead{(J2000.0)} & 
\colhead{(J2000.0)} & 
\colhead{} & 
\colhead{} & 
\colhead{} &
\colhead{(mJy)} & 
\colhead{(mJy)} & 
\colhead{(arcsec)} & 
\colhead{} &
\colhead{} &
\colhead{}
}
\startdata

J0100-0200 & 01 00 32.22 & -02 00 46.3 & 0.227   & 18.2\tablenotemark{a} & J010032.2-020046 & 5.95 & 0.152 & 0.1 & 1.39 & \nodata & Quasar \\
J0103+0037 & 01 03 52.47 & +00 37 39.9 & 0.704   & 17.54  & J010352.4+003739 & 2.22   & 0.143 & 0.2 & 0.63 & \nodata & Quasar \\
J1022-0113 & 10 22 43.73 & -01 13  2.1 &   ?     & 16.54  & J102243.7-011302 & 37.96  & 0.183 & 0.2 & 1.46\tablenotemark{b} & QSO? & BL Lac \\
J1032+0202 & 10 32 42.44 & +02 02 28.9 & 0.537   & 18.67  & J103242.5+020229 & 1.73   & 0.148 & 1.4 & 1.00 & QSO? & Quasar \\
J1143+0113 & 11 43 54.02 & +01 13 43.3 & 1.282   & 17.54  & J114354.0+011343 & 0.80   & 0.162 & 0.5 & 0.12 & QSO? & Quasar \\
J1250+0158 & 12 50 06.82 & +01 58 03.8 & 0.941   & 18.45  & J125006.8+015803 & 155.93 & 0.153 & 0.7 & 2.81 & QSO? & Quasar \\
J1316-0036 & 13 16 37.25 & -00 36 35.8 & 0.895   & 18.37  & J131637.2-003635 & 1.94   & 0.135 & 0.3 & 0.88 & QSO?? & Quasar \\
J1333+0012 & 13 33 56.01 & +00 12 29.1 & 0.918   & 17.64  & J133356.0+001229 & 2.07   & 0.141 & 0.9 & 0.61 & QSO? & Quasar \\
J1334+0102 & 13 34 50.44 & +01 02 18.7 & 0.246   & 18.2\tablenotemark{a} & J133450.4+010220 & 1.70   & 0.140 & 1.5 & 0.85 & \nodata & Quasar \\

\enddata

\tablenotetext{a}{magnitude estimated using peak intensity from APM scan of UKST plate.}
\tablenotetext{b}{BL Lac candidate without confirmed redshift. Redshift z=0.75 adopted in calculation of $LogR^*$.}

\end{deluxetable}

\begin{deluxetable}{lrr}
\tabletypesize{\footnotesize}
\tablewidth{0pt}
\tablenum{6}
\tablecaption{Summary of LBQS/FIRST Cross-Correlation}

\tablehead{
\colhead{} &
\colhead{\bf Stellar} &
\colhead{\bf Non-Stellar}
}
\startdata
Number of LBQS source catalogue objects: & $\sim100000$ & $\sim40000$ \\
Cross-correlation matches with $\sim25000$ FIRST sources: & 101 & 495 \\
Unprocessable objective prism spectra: & 18 & 115 \\
LBQS/FIRST matches: & 83 & 380 \\
Eliminated due to low peak surface brightness: & & 222 \\
Observed in LBQS (through Paper VI): & 64 & 9 \\
Unobserved matches: & 19 & 149 \\
Predicted number of chance coincidences: & $\simeq 10$ & $\simeq 4$ \\
& & \\
{\bf New Observations} & & \\
Keck Spectra: & 11 & 35 \\
MMT Spectra: & 1 & 26 \\
Becker Spectra: & 7 & \\
2dfGRS Spectra: & & 31 \\
NED Identifications: & & 7\\
Total : & 19 & 99 \\
& & \\
{\bf New Quasars:} & 7\tablenotemark{a} & 2 \\
& & \\

\enddata
\tablenotetext{a}{6 quasars + 1 BL Lac (previously observed but misclassified).}
\end{deluxetable}

\clearpage

\begin{deluxetable}{cc}
\tabletypesize{\footnotesize}
\tablewidth{0pt}
\tablenum{7}
\tablecaption{Adopted LBQS Completeness}

\tablehead{
\colhead{\bf Redshift} &
\colhead{\bf Fractional Completeness}
}
\startdata
0.20 & 0.840\\
0.30 - 0.65 & 0.920\\
0.70 & 0.865\\
0.75 - 0.85 & 0.840\\
0.90 & 0.890\\
0.95 - 3.00 & 0.920\\
\enddata
\end{deluxetable}

%\clearpage

\begin{deluxetable}{crcc}
\tabletypesize{\footnotesize}
\tablewidth{0pt}
\tablenum{8}
\tablecaption{Fraction of Radio--detected Quasars}

\tablehead{
\colhead{\bf Magnitude Range} &
\colhead{\bf Total} &
\colhead{\bf Radio--detected} &
\colhead{\bf Fraction}
}
\startdata
16.00 - 17.50 & 61 & 12 & 0.20\\
17.51 - 18.00 & 114 & 20 & 0.18\\
18.01 - 18.25 & 104 & 14 & 0.13\\
18.26 - 18.50 & 157 & 16 & 0.10\\
18.51 - 18.80 & 143 & 14 & 0.10\\
\enddata
\end{deluxetable}

%\clearpage

\begin{deluxetable}{ccccc}
\tabletypesize{\scriptsize}
\tablewidth{0pt}
\tablenum{9}
\tablecaption{FBQS and LBQS Composite Colours}

\tablehead{
\colhead{Redshift} &
\multicolumn{2}{c}{FBQS Composite} &
\multicolumn{2}{c}{LBQS Composite} 
\\
\colhead{} &
\colhead{$B_J-E$} &
\colhead{$O-E$} &
\colhead{$B_J-E$} &
\colhead{$O-E$}
}
\startdata
0.20 & 0.41 & 0.27 & 0.28 & 0.13\\
0.25 & 0.52 & 0.36 & 0.36 & 0.18\\
0.30 & 0.66 & 0.50 & 0.45 & 0.28\\
0.35 & 0.49 & 0.35 & 0.36 & 0.21\\
0.40 & 0.36 & 0.24 & 0.27 & 0.14\\
0.50 & 0.31 & 0.21 & 0.24 & 0.14\\
0.60 & 0.27 & 0.20 & 0.22 & 0.14\\
0.70 & 0.35 & 0.30 & 0.32 & 0.26\\
0.80 & 0.40 & 0.37 & 0.39 & 0.34\\
0.90 & 0.49 & 0.44 & 0.46 & 0.41\\
1.00 & 0.54 & 0.47 & 0.51 & 0.43\\
1.10 & 0.52 & 0.46 & 0.47 & 0.39\\
1.20 & 0.62 & 0.52 & 0.56 & 0.45\\
1.30 & 0.72 & 0.59 & 0.67 & 0.53\\
1.40 & 0.58 & 0.45 & 0.53 & 0.39\\
1.50 & 0.52 & 0.41 & 0.47 & 0.36\\
1.60 & 0.51 & 0.42 & 0.47 & 0.38\\
1.70 & 0.46 & 0.38 & 0.42 & 0.35\\
1.80 & 0.43 & 0.29 & 0.38 & 0.27\\
1.90 & 0.40 & 0.21 & 0.36 & 0.20\\
2.00 & 0.38 &  & 0.36 & \\
2.10 & 0.34 &  & 0.35 & \\
2.20 & 0.30 &  & 0.34 & \\
2.30 & 0.33 &  & 0.40 & \\
2.40 & 0.39 &  & 0.48 & \\
2.50 & 0.38 &  & 0.47 & \\
\enddata
\end{deluxetable}

\clearpage

\center{\bf FIGURE CAPTIONS}

\figcaption{Spectra of 9 (of 12 in total) new LBQS quasars not
included in Paper VI.  Most observations were taken through narrow
slits so the units on the flux scale are arbitrary.}

\figcaption{Spectra of the 9 additional LBQS--FIRST objects.
J1022--0113 is a BL Lac object observed as part of the LBQS and
mis--identified as a star. The remaining 8 objects were discovered
from the cross--correlation of the LBQS and FIRST source catalogues.}

\figcaption{Spectrum of J1438+0032 ($z=0.209$). The narrow
emission lines and the presence of stellar absorption features are
consistent with the identification as a Seyfert 2 nucleus.}

\figcaption{Distribution of ${\rm log} R^*$ versus redshift
for the known LBQS--FIRST objects from Table 2 ($\bullet$) and the
9 new LBQS--FIRST objects ($\circ$) from Table 5. The BL Lac object,
J1022--0113, has been plotted at redshift $z=0.75$.}

\figcaption{Distribution of the Half-Power Point (HPP) of the
objective--prism spectra for LBQS quasars as a function of redshift.
The 67 previously known LBQS--FIRST objects ($\bullet$) and the 9
additional LBQS--FIRST objects ($\circ$) are also shown.  The
horizontal dash--dot line at HPP=0 indicates the boundary of the LBQS
colour--selection algorithm; objects faling below the line being
defined as of unusually blue colour. The loci of the LBQS and FBQS
composite spectra are shown as the solid and dashed lines
respectively.}

\figcaption{Redshift distribution of FBQS and LBQS quasars. (a):
distribution of the 565 FBQS quasars with redshifts $0.2 \le z \le 3.0$
and magnitudes $15.5 \le E \le 17.8$ from Table 2 of White et al.
(2000) (solid line).  The distribution of the full LBQS sample,
adopting the redshift completeness factors listed in Table 7, over the
same redshift range is shown as the dotted line. The LBQS histogram has
been normalised to the same total area as the FBQS histogram. (b): the
distribution of the 565 FBQS quasars shown in panel (a) is reproduced
(solid line) together with associated error bars. The redshift
distribution of the LBQS sample limited at $15.5 \le E \le 17.8$ is
shown as the dotted line, together with associated error bars. The LBQS
histogram has been normalised to the same total area as the FBQS
histogram and the x--coordinate of the error bars displaced to slightly
larger redshifts for clarity.}

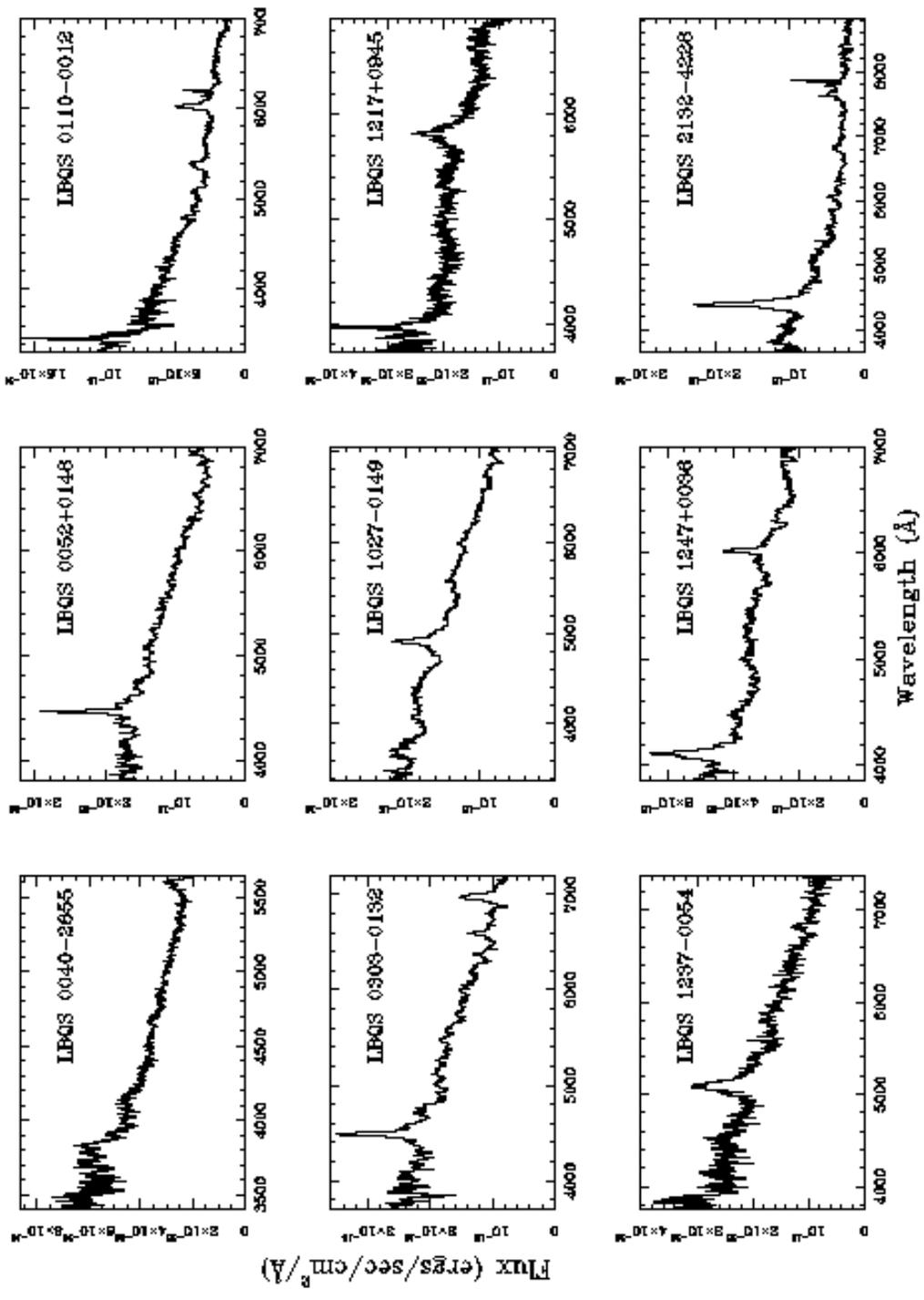
\begin{figure}
\figurenum{1}
\epsscale{1.75}
\plotone{plot1.epsi}
\caption{}
\end{figure}

\clearpage

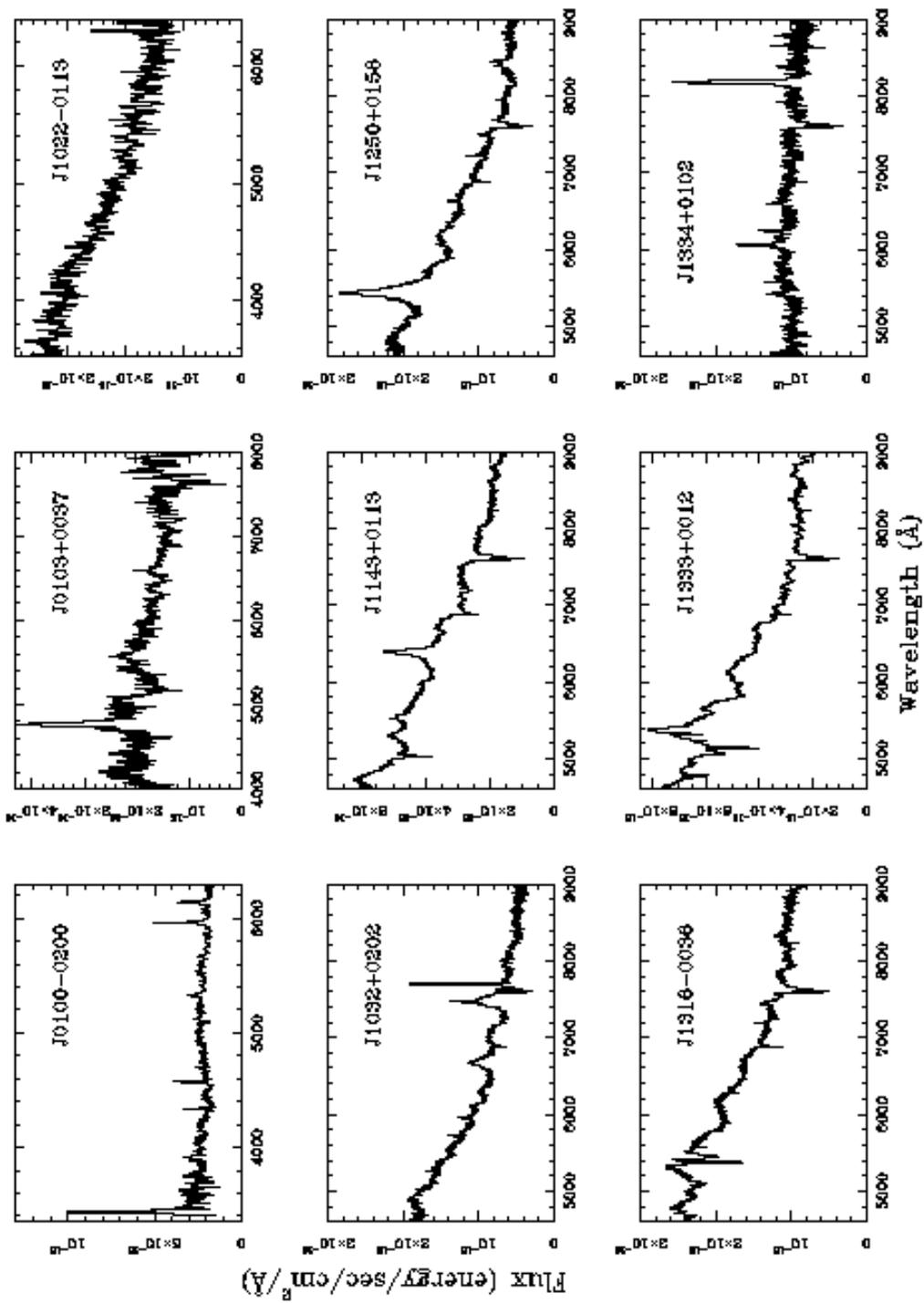
\begin{figure}
\figurenum{2}
\epsscale{1.75}
\plotone{plot2.epsi}
\caption{}
\end{figure}

\clearpage

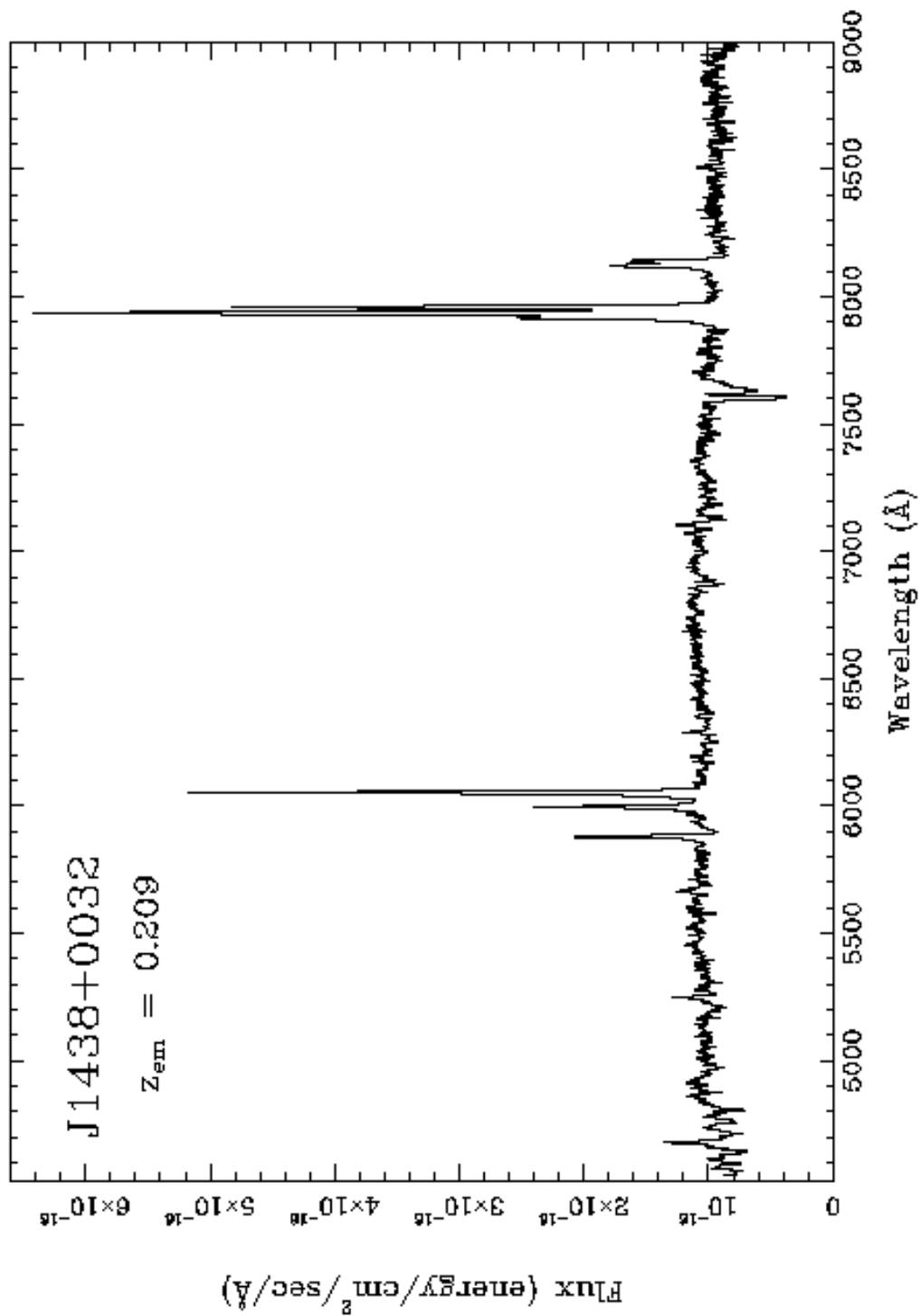
\begin{figure}
\figurenum{3}
\epsscale{1.75}
\plotone{plot3.epsi}
\caption{}
\end{figure}

\clearpage

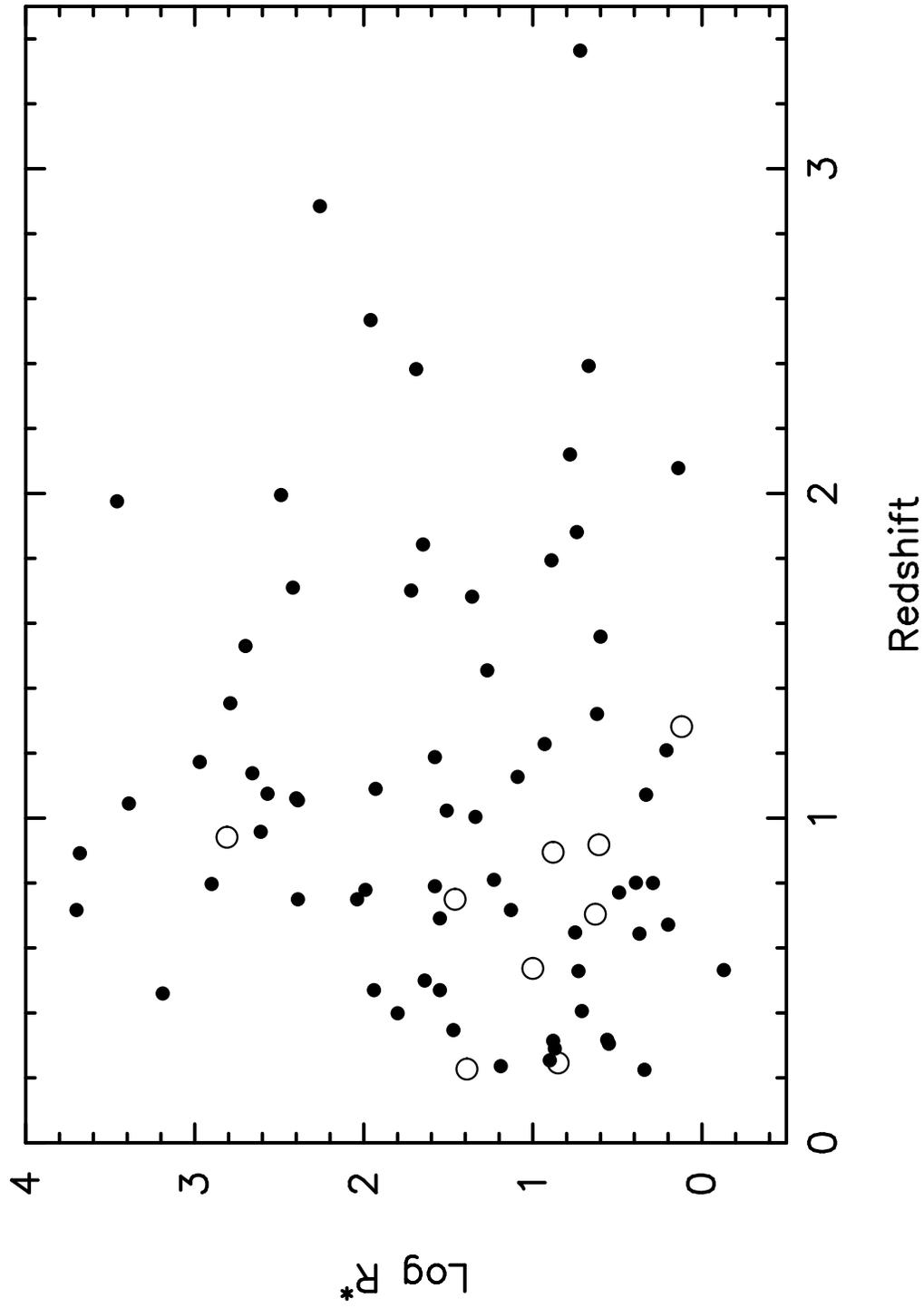
\begin{figure}
\figurenum{4}
\epsscale{1.75}
\plotone{plot4.epsi}
\caption{}
\end{figure}

\clearpage

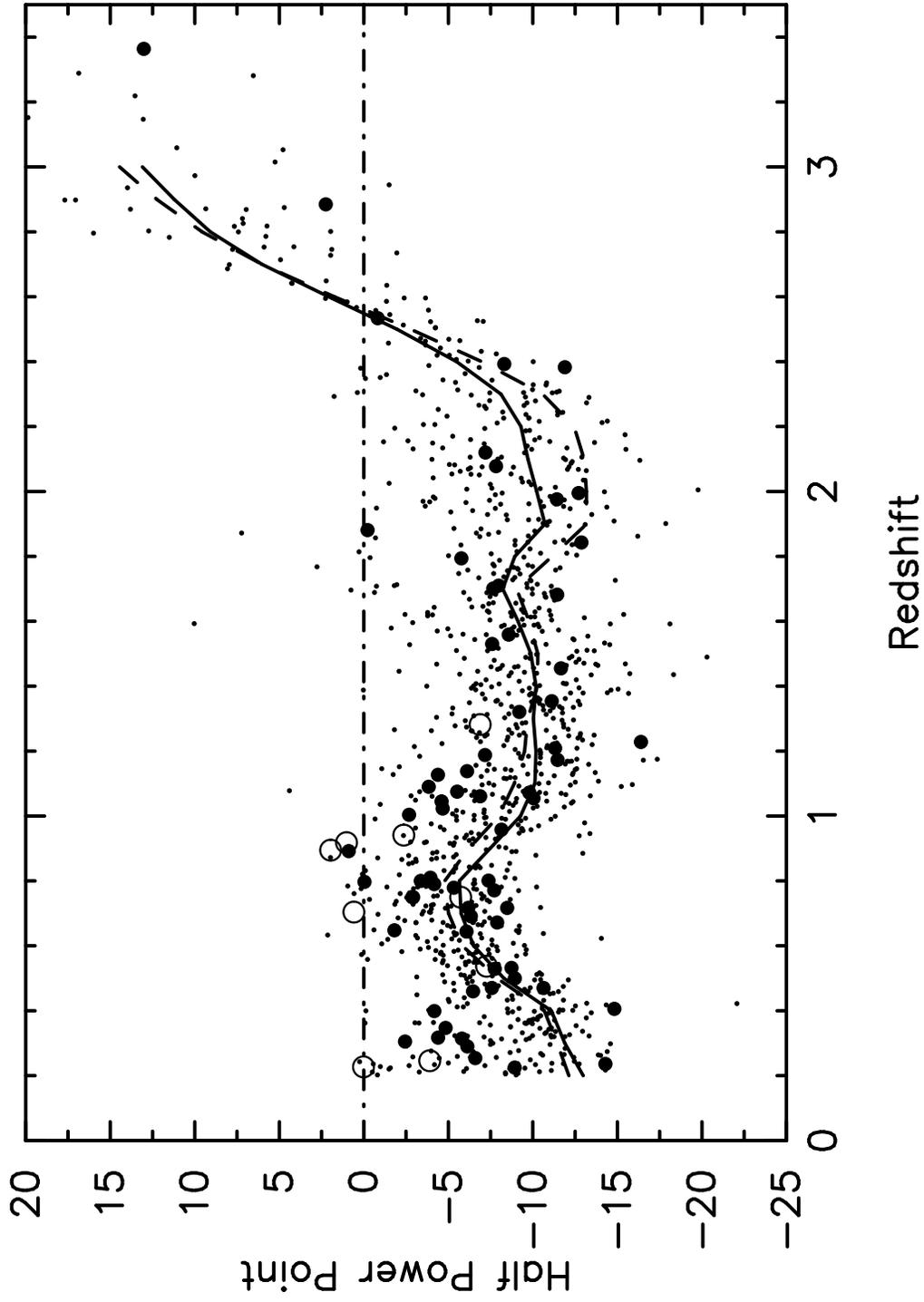
\begin{figure}
\figurenum{5}
\epsscale{1.75}
\plotone{plot5.epsi}
\caption{}
\end{figure}

\clearpage

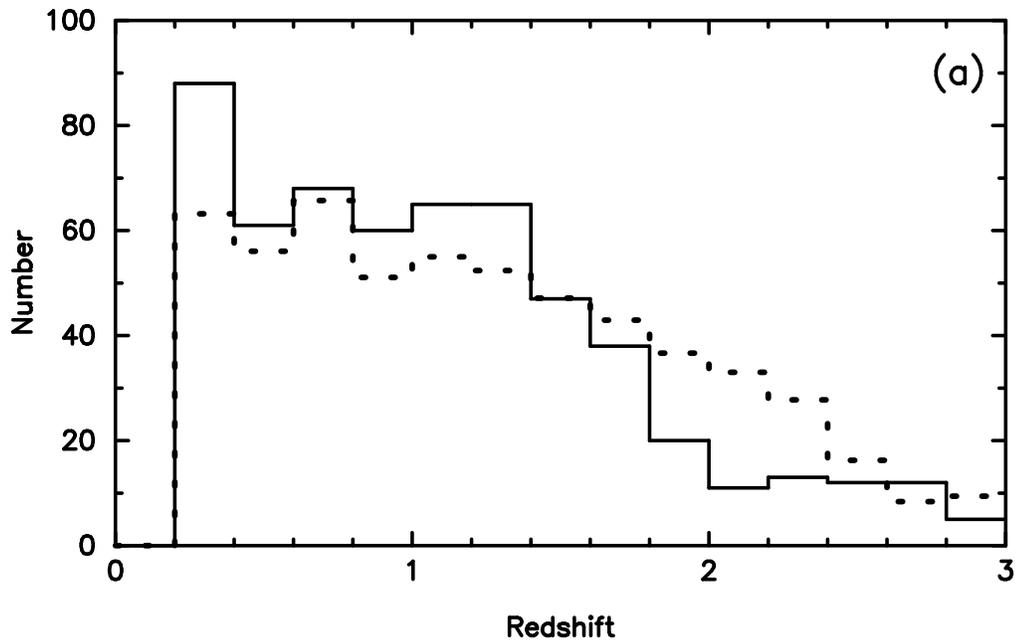
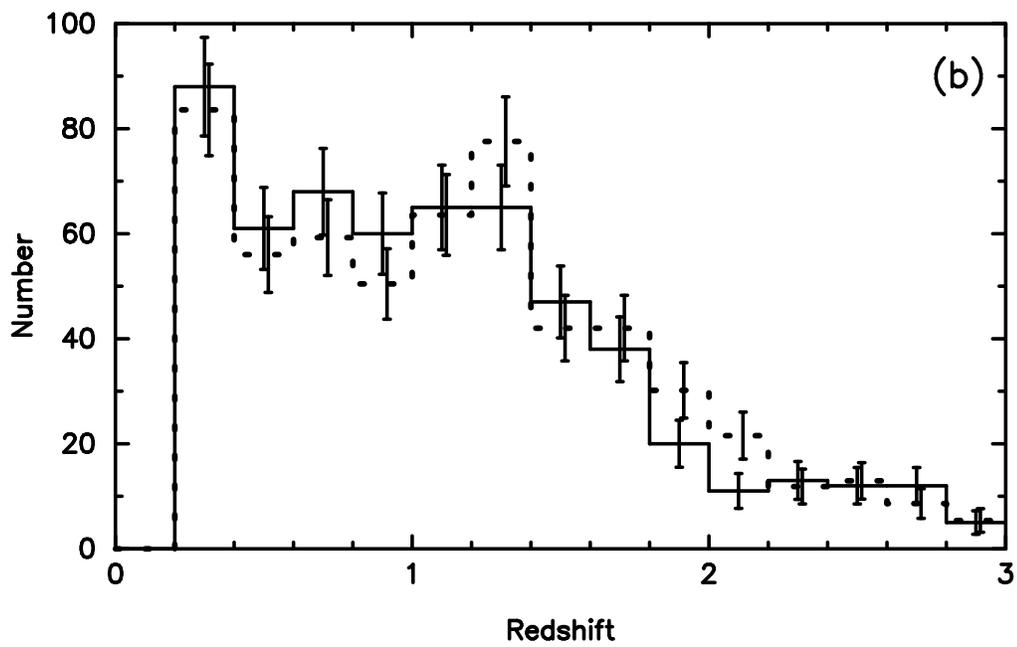
\begin{figure}
\figurenum{6}
\epsscale{1.75}
\plotone{plot6.epsi}
\caption{}
\end{figure}


\begin{thebibliography}{}

\bibitem[Appenzeller et al. (1998)]{ap98}Appenzeller, I., et al. 1998, ApJS,
117, 319
\bibitem[Becker et al. 2000]{be00} Becker, R. H., White, 
R. L., Gregg, M. D., Brotherton, M. S., Laurent-Muehleisen, S. A., \& 
Arav, N. 2000, \apj, 538, 72 
\bibitem[Becker, White \& Helfand (1995)]{wh95} Becker, R. H., White, R. L., \& Helfand, D. J. 1995, \apj, 450, 559
\bibitem[Boyle et al. (1990)]{boyle90} Boyle, B. J., Fong, R., Shamks, T., \&
Peterson, B. A. 1990, \mnras, 243, 1
\bibitem[Brotherton et al. (2001)]{br01} Brotherton, M. S., 
Tran, H. D., Becker, R. H., Gregg, M. D., Laurent--Muehleisen, S. A., \& 
White, R. L. 2001, \apj, 546, 775 
\bibitem[Bunclark \& Irwin (1983)]{bun83} Bunclark, P. S., \& Irwin, M. J. 1983,
in Proceedings of the Statistical Methods in Astronomy Symposium, ESA Special
Publication No. SP--201 (ESA, Noordwijk), p. 195
\bibitem[Clements et al. (1996)]{Clem96} Clements, D. L., 
Sutherland, W. J., Saunders, W., Efstathiou, G. P., McMahon, R. G., 
Maddox, S., Lawrence, A., \& Rowan-Robinson, M. 1996, \mnras, 279, 459 
\bibitem[Colless (1999)]{colless99} Colless, M. M. 1999, Phil. Trans. Roy. Soc. Lond. A, 357, 105 
\bibitem[Cristiani et al. (1995)]{crist95} Cristiani, S., et al. 1995, \aaps, 112, 347
\bibitem[Eddington (1913)]{ed13}Eddington, A. S. 1913, \mnras, 73.359
\bibitem[Francis et al. (1991)]{fr91} Francis, P. J., Hewett, P. C., Foltz, C. B., Chaffee, F. H., Weymann, R. J., \& Morris, S. L. 1991, \apj, 373, 465
\bibitem[Goldschmidt 1993]{go93}Goldschmidt, P. 1993, Ph.D. thesis, University of Edinburgh
\bibitem[Hawkins and Veron (1995)]{hv95} Hawkins, M. R. S., \& V\'{e}ron, P. 1995,
\mnras, 275, 1102
\bibitem[Hewett et al. (1993)]{he93} Hewett, P. C., Foltz, C. B., \& Chaffee, F. H. 1995, \apj, 406, L43
\bibitem[Hewett et al. (1995)]{he95} Hewett, P. C., Foltz, C. B., \& Chaffee, F. H. 1995, \aj, 109, 1498
\bibitem[H{\o}g et al.(2000)]{ho2000} H{\o}g, E., Fabricius, 
C., Makarov, V. V., Bastian, U., Schwekendiek, P., Wicenec, A., Urban, S., 
Corbin, T., \& Wycoff, G. 2000, \aap, 357, 367
\bibitem[Hooper et al. (1995)]{hoo95} Hooper, E. J., Impey, C. D., Foltz, C. B.,
\& Hewett, P. C. 1995, \apj, 445, 62
\bibitem[Minkowski \& Abell (1963)]{min63} Minkowski, R.L., \& Abell, G.O.
1963, Stars and Stellar Systems Vol III, Basic Astronomical Data,
University of Chicago Press, 481
\bibitem[Oke et al. (1995)]{oke95}Oke, J. B., Cohen, J. G., Carr, M., Cromer, J., Dingizian, A., Harris, F. H., Labrecque, S., Lucinio, R., Schall, W., Epps H., and Miller, J. 1995, \pasp, 107, 375 
\bibitem{ros88} Roser, S., \& Bastian, U. 1988 A\&AS, 74, 449
\bibitem[Shectman et al.(1996)]{sch96} Shectman, S. A., 
Landy, S. D., Oemler, A., Tucker, D. L., Lin, H., Kirshner, R. P., \& 
Schechter, P. L. 1996, \apj, 470, 172 
\bibitem[Schneider et al. (1994)]{ssg94} Schneider, D. P., Schmidt, M., \&
Gunn, J. E. 1994, \aj, 107, 1245
\bibitem[Stocke et al.(1991)]{st91} Stocke, J. T., Morris, 
S. L., Gioia, I. M., Maccacaro, T., Schild, R., Wolter, A., Fleming, T. 
A., \& Henry, J. P. 1991, \apjs, 76, 813 
\bibitem[Stocke et al. (1992)]{st92}Stocke, J. T., Morris, S. L., 
Weymann, R. J., \& Foltz, C. B. 1992, \apj, 396, 487
\bibitem[Suntzeff et al. (1996)]{sun96} Suntzeff, N., et al. 1996, IAU Circ., 6490, 1
\bibitem[Usher \& Mitchell (2000)]{ush00}Usher, P. D., \& Mitchell, K. J.
2000, \aj, 120, 1683
\bibitem[Voges et al. (1999)]{vo99}Voges, W., et al. 1999, A\&A, 349, 389
\bibitem[Warren et al. (1991)]{warren91} Warren, S. J., Hewett, P. C., \& Osmer,
P. S. 1991, \apjs, 76, 23
\bibitem[White et al. (2000)]{wh2000} White, R. L. et al. 2000, \apjs, 126, 123
\bibitem[White et al. (1997)]{wh1997} White, R. L., Becker, R. H., Helfand, D.J., \& Gregg, M. D. 1997, \apj, 475, 479
\bibitem[Wills and Wills (1974)]{wills74} Wills, D., \& Wills, B.J. 1974, \apj, 190, 271
\end{thebibliography}
\end{document}